\documentclass[10pt, journal, a4paper, twocolumn, final]{IEEEtran}
\usepackage[T1]{fontenc}
\usepackage[active]{srcltx}
\usepackage{array}
\usepackage{float}
\usepackage{amsthm}
\usepackage{amsmath}
\usepackage{amssymb}
\usepackage{graphicx}
\usepackage{setspace}
\usepackage{changebar}
\usepackage{color}
\usepackage{booktabs}
\usepackage{colortbl}
\usepackage{bbding}
\usepackage{algorithmicx}
\usepackage{algpseudocode}

\makeatletter

\providecommand{\tabularnewline}{\\}
\floatstyle{ruled}
\newfloat{algorithm}{tbp}{loa}
\providecommand{\algorithmname}{Algorithm}
\floatname{algorithm}{\protect\algorithmname}

\theoremstyle{plain}
\newtheorem{thm}{\protect\theoremname}
\theoremstyle{remark}
\newtheorem{rem}[thm]{\protect\remarkname}

\@ifundefined{date}{}{\date{}}
\IEEEoverridecommandlockouts
\usepackage{ifpdf}
\usepackage{cite}
\ifCLASSOPTIONcompsoc
\usepackage[caption=false,font=normalsize,labelfont=sf,textfont=sf]{subfig}
\else
\usepackage[caption=false,font=footnotesize]{subfig}
\fi
\usepackage{amssymb}
\usepackage{amsmath}

\@ifundefined{showcaptionsetup}{}{%
 \PassOptionsToPackage{caption=false}{subfig}}
\usepackage{subfig}
\makeatother

\providecommand{\remarkname}{Remark}
\providecommand{\theoremname}{Theorem}

\begin{document}

\title{Space-Time Hierarchical-Graph Based Cooperative Localization in Wireless Sensor Networks }
\author{Tiejun Lv, \emph{Senior Member, IEEE}, Hui Gao, \emph{Member, IEEE}, Xiaopeng Li, \\
Shaoshi Yang, {\em Member,~IEEE}, and Lajos Hanzo, {\em Fellow,~IEEE}%
\thanks{Copyright (c) 2015 IEEE. Personal use of this material is permitted. However, permission to use this material for any other purposes must be obtained from the IEEE by sending a request to pubs-permissions@ieee.org.

This work is financially supported by the National Natural Science
Foundation of China (NSFC) (Grant No. 61271188) and the Fundamental
Research Funds for the Central Universities (Grant No. 2014RC0106).%

T. Lv, H. Gao, and X. Li are with the School of Information and Communication
Engineering, Beijing University of Posts and Telecommunications, Beijing,
China 100876 (e-mail: \{lvtiejun, huigao, lixiaopeng\}@bupt.edu.cn).%

S. Yang and L. Hanzo are with the School of Electronics and
Computer Science, University of Southampton, Southampton, SO17 1BJ, U.K. (email:
\{sy7g09, lh\}@ecs.soton.ac.uk).}
}

\markboth{Accepted to appear on IEEE Transactions on Signal Processing, Sept. 2015}%
{Shell \MakeLowercase{\textit{et al.}}: Bare Demo of IEEEtran.cls
for Journals}

\maketitle
\begin{abstract}
It has been shown that cooperative localization is capable of improving both the positioning accuracy and coverage in scenarios where the global positioning system (GPS) has a poor performance. However, due to its potentially excessive computational complexity, at the time of writing the application of cooperative localization remains limited in practice.
In this paper, we address the efficient cooperative positioning problem in wireless sensor
networks. A space-time hierarchical-graph based scheme exhibiting fast convergence is proposed for localizing the agent nodes.
In contrast to conventional methods, agent nodes are divided into different layers with the aid of the space-time hierarchical-model  and their positions are estimated gradually.
In particular, an information propagation rule is conceived upon considering the quality of positional information.
According to the rule, the information always propagates from the upper layers
to a certain lower layer and the message passing process is further optimized at each layer. Hence, the potential error propagation can be mitigated.
Additionally, both position estimation and position broadcasting are carried out by the sensor nodes. Furthermore, a sensor activation mechanism is conceived, which is capable of significantly reducing both the energy consumption and the network traffic overhead incurred by the localization process.
The analytical and numerical results provided demonstrate the superiority of our space-time hierarchical-graph based cooperative localization scheme over the benchmarking schemes considered.
\end{abstract}

\section{Introduction}

\renewcommand{\figurename}{Fig.}\renewcommand{\tablename}{TABLE} Having accurate positional
information is a vital requirement in various applications, such as
traffic control, security tracking, search and rescue operations,
medical assistance and the Internet of Things \cite{winnetwork2011,patwari2005locating,gezici2005localization,ditarantolocationaware2014,shenfundamental2010,shensensor2014,jennings1997cooperative}.
This wide range of applications has inspired growing interests in
localization in wireless sensor networks (WSNs)\cite{zhang2009uwb,shen2012network, masazadedynamic2012}.
In a WSN, it is usually assumed that anchor nodes whose positions are known can be employed for locating the agent nodes.

In general, the existing positioning techniques may be classified
into two categories: non-cooperative methods and cooperative methods.
For the non-cooperative methods \cite{dardari2009ranging,guvenc2006toa,xu2008delay,youssef2010new},
an agent node is localized by measuring the distances between itself
and the anchors, whose positions are known. In contrast to the non-cooperative
methods, cooperation among the agents is capable of improving positioning accuracy,
wherein an agent measures the distances from itself to both the
anchors and the other agents \cite{wymeersch2009cooperative}.
Recently, cooperative localization has attracted significant research interests \cite{wymeersch2009cooperative,gholami2011robust,gholami2011wireless,ihler2005nonparametric,pedersen2011variational,savic2009sensor,savic2011optimized,das2010censored,das2012censoring,denis2009scheduling,savarese2001location,vaze2012bounds}.
Most of the state-of-the-art cooperative localization methods are based on the \emph{point estimation} philosophy, as exemplified by the maximum likelihood method\cite{wymeersch2009cooperative},
the non-linear least squares method\cite{wymeersch2009cooperative}, the two-step linear least squares method\cite{gholami2011robust} and
the geometric optimization methods\cite{gholami2011wireless}. However,
these approaches estimate the positions of all the agent nodes
in a ``hard-decision'' manner, and they do not take the statistical properties
of the estimation error into further consideration.
Therefore, these schemes suffer from the lack of a statistical interpretation
that could be useful for developing new methods exhibiting an improved performance.

Compared to the above-mentioned \emph{point estimation} methods, Bayesian approaches
consider the position of an agent node as a random variable, and their
goal is to estimate the \textit{a posteriori} marginal probability density functions
(PDF) of all agents' positions, which is regarded as a localization
inference problem. Nevertheless, the marginalization
of the joint \textit{a posteriori} PDF of all the agent nodes imposes a potentially excessive computational complexity, especially in large-scale networks. As a remedy, graphical model based methods have been
proposed in\cite{kschischang2001factor,wymeersch2009cooperative}
for simplifying the localization inference problem\cite{ihler2005nonparametric,savic2009sensor,wymeersch2009cooperative}.
The message-passing methods constitute the class of ``best-known'' methods which
make inference on the graph, resulting in an approximately optimal
solution\cite{wymeersch2009cooperative,kschischang2001factor}. In particular, the belief propagation (BP) algorithm
\cite{pearl_probabilistic_1988,savarese2001location} represents an
efficient message-passing method of calculating the marginal distribution
based on probabilistic graphical models, and it can be applied to cooperative
localization in WSNs\cite{wymeersch2009cooperative,sudderth2010nonparametric}.
It is noted that the computational complexity of the BP algorithm is proportional
to the total number of links on the graph, thus it is much more efficient
than the naive ``brute-force'' method, which imposes an exponentially increasing computational complexity when computing the
marginal probabilities. However, the standard BP is unable to resolve a non-Gaussian uncertainty, which is often encountered in a practical localization scenario.
In contrast to the standard BP, the nonparametric belief propagation (NBP) is a sample-based algorithm, which can be readily applied in non-Gaussian inference problems\cite{sudderth2010nonparametric,ihler2005nonparametric}.
Therefore, the NBP is more suitable for localization. Having said this, an impediment
of NBP (and also of BP) is that for graphs exhibiting loops, there
are no guarantees concerning the quality of the marginal beliefs as well as the convergence
of the algorithm. Moreover, the employment of NBP is impractical in dense WSNs, because
directly employing all the links of the graph for localization without any pre-selection leads to
a high computational complexity, a high network traffic, as well as an increased energy
consumption. In order to mitigate the influence of loops in the graph, a generalized
belief propagation (GBP) algorithm was proposed in \cite{yedidia2003understanding,savic2009sensor}.
However, GBP also remains unsuitable for large-scale WSNs owing to its excessive computational
complexity. In addition, although the NBP defined over spanning trees with breadth first search (NBP-BFS)\cite{vladimir2010indoor}, the tree-reweighted belief propagation (TRW-BP) \cite{savic2011optimized} and the NBP defined over the minimum spanning tree (NBP-MIN)\cite{Xiaopeng_2015:NBP_MIN} techniques are potentially capable of outperforming NBP when dealing with loops, they also
have their particular limitations. To elaborate a little further, NBP-BFS exhibits poor performance in WSNs having sparse nodes, NBP-MIN is less attractive than the standard NBP in WSNs having short transmission radius, while TRW-BP is unsuitable for a distributed implementation,
since it requires the knowledge of the PDFs over all spanning trees.

On the other hand, in order to facilitate a more practical implementation that imposes both a reduced network traffic
and a low computational complexity, a technique referred to as \textit{information censoring} has been considered in \cite{denis2009scheduling,das2010censored,das2012censoring},
which results in carefully controlled information propagation. In \cite{das2010censored,das2012censoring},
the Cram\'{e}r-Rao bound (CRB) was considered as a criterion for removing any unreliable positional information. Nevertheless, it is possible that some censored information
remains helpful to particular agent nodes. Therefore, a more judicious control of information propagation (message passing) is
required for improving the positioning quality and for reducing the computational complexity.
The bootstrap percolation approach of \cite{adler1991bootstrap,sarkar2013distributed}
is capable of controlling error propagation, hence it has attracted substantial attention in
recent years. It was shown in \cite{vaze2012bounds} that bootstrap
percolation relying on a hard threshold is capable of carrying out iterative localization\cite{savarese2001location}.
However, some agent nodes that do not satisfy the above hard threshold cannot
be localized with the aid of this approach. In our previous work \cite{cai2013}, we have shown that for non-Bayesian
cooperative positioning (e.g. point estimation), soft threshold aided bootstrap percolation is capable of improving the localization performance, despite  reducing both the computational complexity and the energy
consumption.

In this paper, we aim to extend the bootstrap percolation approach for solving the Bayesian localization problem. To achieve this goal, a novel NBP-based cooperative localization scheme is
proposed, which is capable of reducing the network traffic, whilst accelerating the
convergence of the positioning algorithm. More specifically, a space-time hierarchical-graph based description of the network is proposed, which relies on the bootstrap percolation philosophy for facilitating the implementation of
NBP-based cooperative localization, where the agent nodes are divided into different
layers, and the localization process is carried out in
a layer-by-layer manner. In particular, based on the framework of
the space-time hierarchical-graph description of the network, a beneficial information propagation rule
is conceived for controlling the information propagation. This rule includes
a joint design of inter-layer message passing and intra-layer message
passing. Among different layers, the messages always propagate from the upper
layers to a certain lower layer. However, the messages originating from lower layers, which contain
the erroneous information, are not allowed to propagate to the upper layers. As a result, the impact of any erroneous information can be mitigated. Within a single layer, the messages are allowed to be exchanged among intra-layer sensor nodes, when
there are insufficient reference nodes (including both agent and anchor
nodes) in connection. By employing the proposed information propagation
rule, the positioning accuracy can be significantly improved. Additionally,
in order to manage the energy consumption and network traffic more effectively, both position estimation
and position broadcasting are supported by the sensor nodes.
In contrast to traditional cooperative methods, we put our emphasis on the behavior
constraints of sensor nodes by employing a specific activation mechanism.
Note that the activated agent nodes are capable of estimating their positions
and of broadcasting their positional information, whereas the inactive agent
nodes remain silent in order to save energy and to reduce the network traffic.
Our main contributions in this paper are summarized as follows.

\begin{table*}[t]
\centering{}\caption{List of Major Variables/Notation Used in This Paper}
\begin{small}
\begin{tabular}{|>{\raggedright}p{3.5cm}|>{\raggedright}p{8cm}|}

\hline
Variable/Notation  & Definition\tabularnewline
\hline
$\mathbf{x}_{i}$ & Two-dimensional position of agent $i$, $i\in I\,:=[1,\ldots,N]$\tabularnewline
$\mathbf{x}_{j}$ & Two-dimensional position of anchor $j$, $j\in J\,:=[N+1,\ldots,N+M]$\tabularnewline
$\tilde{d}_{vi}$  & Noisy measurement of the distance between node $v$ and node $i$ \tabularnewline
$d(\mathbf{x}_{v},\mathbf{x}_{i})$ & Actual distance between node $v$ and node $i$\tabularnewline
$n_{vi}$ & Noise of the distance measurement between node $v$ and node $i$\tabularnewline
$R$ & Transmission radius\tabularnewline
$S_{i}$ & Set of neighboring nodes of node $i$\tabularnewline
$p(\mathbf{x}_{i})$ & Prior distribution of the position of node $i$\tabularnewline
$Q$ & A clique which is a subset of $V$\tabularnewline
$\psi_{i}(\mathbf{x}_{i})$ & The single-node potential of node $i$\tabularnewline
$\psi_{vi}(\mathbf{x}_{v},\mathbf{x}_{i})$  & The pairwise potential between node $i$ and $v$\tabularnewline
$b_{i}^{t}(\mathbf{x}_{i})$ & Belief (or marginal distribution) on the position of node $i$ at time slot $t$\tabularnewline
$m_{vi}^{t}(\mathbf{x}_{i})$  & Message from node $v$ to node $i$ at time slot $t$\tabularnewline
$\gamma(\mathbf{x}_{i})$ & Redundancy at node $i$ to represent the diversity of two arbitrary distributions\tabularnewline
$\mathcal{F}_{i}$  & A set containing the selected sensor nodes for
information fusion, hence it is a subset of $S_{i}$.\tabularnewline
$\mathcal{A}_{l}$  & The set of agents that are activated on layer $l$.\tabularnewline
$\mathcal{R}^{l}$ & The set of candidate reference nodes on layer $l$\tabularnewline
$L$ & The number of layers in the hierarchical NBP algorithm\tabularnewline
$K$ & The number of weighted samples of NBP\tabularnewline
\hline
\end{tabular}
\end{small}
\end{table*}

\newcounter{numcount2} \begin{list}{ \arabic{numcount2})}{\usecounter{numcount2} \setlength{\itemindent}{-1em}\setlength{\rightmargin}{0em}} \setlength\leftskip{1ex}
\item{An NBP-based cooperative localization scheme, which exploits the proposed space-time hierarchical-graph, is conceived for WSNs. By formulating the localization inference problem subject to additional constraints, a reliable solution can be guaranteed. Provided that we stipulate the constraint that there should be at least three reference nodes in support of position estimation, reliable positional information may be gleaned. Additionally, this constraint ensures that the feasible region is  shrunken into a small region surrounding the optimal solution.}
\item{Inspired by the bootstrap percolation strategy, a space-time hierarchical-graph is formulated, which is capable of mitigating the influence of loops contained in the graph. The soft connection constraint invoked by our strategy ensures that most of the agent nodes share a sufficiently high number of reference nodes for their position estimation action. Compared to the traditional spanning tree method of \cite{vladimir2010indoor} that does not rely on a sufficiently high number of reference nodes, the proposed space-time hierarchical-graph is characterized by multiple layers emerging from invoking the bootstrap percolation strategy. Therefore, the positioning result becomes more accurate.}
\item{An information propagation rule designed for our space-time hierarchical-graph is investigated, which relies on the joint design of the inter-layer as well as intra-layer message passing. The inter-layer message passing is capable of mitigating the error propagation, while the intra-layer  message passing ensures that all beneficial information can indeed be fully exploited. Additionally, with the aid of a beneficial activation mechanism, both efficient information control and energy/traffic management can be achieved by relying on this rule.}
\item{The computational complexity and network traffic of the proposed scheme is analyzed. Since both of them are related to the number of actively participating links of the graph, we adopt the average number of links used in the entire network as the metric of comparing the complexity of different NBP algorithms. We demonstrate that the proposed scheme is capable of significantly reducing the computational complexity and network traffic by censoring the unreliable messages emerging from the lower layers.}
\end{list} The rest of this paper is organized as follows. In Section II, we
describe the system model and formulate the problem considered. In Section III,
we characterize the localization inference problem on a graphical model
and propose a novel space-time hierarchical-graph based scheme for
cooperative localization. We present both the computational complexity and network traffic
analysis of the proposed algorithm in Section IV, and characterize the positioning performance in Section V. Finally, our conclusions are offered
in Section VI.

\textit{Notations}: $\mathcal{N}(\mu,\sigma^{2})$ denotes a normal distribution
with expectation of $\mu$ and standard deviation of $\sigma$;
$|\cdot|$ denotes the cardinality of a set; $||\cdot||$ represents the L2-norm
of a vector; $\mathbb{E}\{\cdot\}$ stands for the expectation operation; $\varnothing$
denotes an empty set; $\mathcal{A}\setminus\mathcal{B}=\{x|x\in\mathcal{A} \; \text{and} \;   x\notin\mathcal{B}\}$
denotes the relative complement of $\mathcal{B}$ in $\mathcal{A}$;
$H(\mathbf{x})$ is the entropy of $\mathbf{x}$; $\mathbb{I}\{P\}$
is the indicator function, which has a value of one when $P$ is true and zero
otherwise; $\text{Cov}[\cdot]$ denotes the covariance; and $\mathbf{I}$ is
the identity matrix. Table I summarizes some of the major variables used
in this paper.

\section{System Model}

We consider a two-dimensional WSN consisting of two types of sensor
nodes: $N$ agents and $M$ anchors. The agents have unknown positions
at $\mathbf{x}_{i},\, i\in I\,:=\{1,\ldots,N$\}, and $\mathbf{x}_{i}$
is a two-dimensional random variable representing the position of
agent node $i$. Similarly, the anchors have known positions at $\mathbf{x}_{j},\, j\in J\,:=\{N+1,\ldots,N+M\}$.
Thus, all sensor node indices are included in $V=I\cup J$. We assume
that agent node $i$ acquires a noisy measurement $\tilde{d}_{vi}$, which is the estimated  distance from sensor node $v$  ($v$ can be either an agent or an anchor) to agent node $i$ during each time slot. Then, we have
\begin{equation}
\tilde{d}_{vi}=d(\mathbf{x}_{v},\mathbf{x}_{i})+n_{vi},\, v\in V,\, i\in I,
\end{equation}
where $d(\mathbf{x}_{v},\mathbf{x}_{i})=||\mathbf{x}_{v}-\mathbf{x}_{i}||$
is the actual distance between node $v$ as well as node $i$, and $n_{vi}$
is the measurement noise, which may obey log-normal,
Gaussian or any other appropriate distribution measured in various deployment districts. It is worth noting that NBP is applicable to all the possible distributions, since NBP is a sample-based nonparametric method. Without loss of
generality, we assume that $n_{vi}\sim\mathcal{N}(0,\sigma_{vi}^{2})$\cite{wymeersch2009cooperative}, and the probability of the event that node $v$ and node $i$ can detect each other is given by
\begin{equation}
p_{d}(\mathbf{x}_{v},\mathbf{x}_{i})=\begin{cases}
1, & ||\mathbf{x}_{v}-\mathbf{x}_{i}||\leq R,\\
0, & \mathrm{otherwise},
\end{cases}
\end{equation}
where $R$ represents the transmission radius. Note that the pair of nodes
$v$ and $i$ involved in the detection have been connected with each other via an edge, which is denoted by $(v,i)$,
satisfying $(v,i)\in E$, where $E$ is a set containing all the edges. Then, for
each sensor node $i\in I$, the set of neighboring
nodes connected to node $i$ is given by $S_{i}=\{v|\,||\mathbf{x}_{v}-\mathbf{x}_{i}||\leq R,\, v\in V\}$.
Additionally, we assume that the position of a sensor node $v$ has an \textit{a priori} distribution
of $p(\mathbf{x}_{v})$. For an agent, $p(\mathbf{x}_{v})$ is a uniform distribution over the entire area, while that of an anchor is the Dirac Delta function
at its true position. Furthermore, we assume that a) the positions of the agent nodes
are independent \textit{a priori}, i.e. we have $p(\mathbf{x}_{1},\ldots,\mathbf{x}_{N})=\prod_{i=1}^{N}p(\mathbf{x}_{i})$;
b) the relative positions during arbitrary time slots are \textit{conditionally
independent} and depend only on the two nodes involved, i.e. we have $p(\{\tilde{d}_{vi}\}|\mathbf{x}_{1},\ldots,\mathbf{x}_{N})=\overset{N}{\underset{i=1}{\prod}}\underset{s\in S_{i}}{\prod}p(\tilde{d}_{si}|\mathbf{x}_{i},\mathbf{x}_{s})$,
where $\{\tilde{d}_{vi}\}$ denotes the set of measurements concerning the distance
from sensor node $i$ to sensor node $v$, $i\in I,$ $v\in V$, $i\neq v$.
Then, the joint PDF of the positions of all agents can be characterized as
\begin{eqnarray}
&p(\mathbf{x}_{1},\ldots,\mathbf{x}_{N},\{\tilde{d}_{vi}\})\nonumber \\& =  \underset{(v,i)\in E}{\prod}p_{d}(\mathbf{x}_{v},\mathbf{x}_{i})\underset{(v,i)\in E}{\prod}p(\tilde{d}_{vi}|\mathbf{x}_{v},\mathbf{x}_{i})
\underset{v\in V}{\prod}p(\mathbf{x}_{v}).
\end{eqnarray}
Our objective is to compute (or approximate) the \textit{a posteriori} marginal
PDF $p(\mathbf{x}_{i}|\{\tilde{d}_{vi}\})$ for each agent node $i$.
Then, either the minimum mean square error (MMSE) estimator
or the maximum \emph{a posteriori} (MAP) estimator can be employed
for completing the location estimation process. In this paper, the MMSE estimator that has a relatively low computational complexity
is invoked, since ideally the position inference process should impose a low computational complexity and a low additional network traffic, which is also the objective of our proposed scheme to be presented in the forthcoming sections.

\section{The Proposed Space-Time Hierarchical Localization Scheme}

In this section, we first present the statistical framework invoked for sensor localization relying on a graph. By formulating the localization inference problem subject to additional constraints, finding an accurate solution is guaranteed by the proposed scheme. Then, upon exploiting the idea of bootstrap percolation subject to soft connection constraints, a space-time hierarchical graphical model is formulated. Based on this model, we conceive an information propagation rule defined over the space-time hierarchical-graph, which requires a joint design of the inter-layer message-passing and intra-layer message-passing. Finally, a new NBP-based localization algorithm is conceived, which carries out the localization inference based on the proposed space-time hierarchical model.

\subsection{Inference on Graph}
\begin{figure}[t]
\begin{centering}
\subfloat[ ]{\begin{centering}
\includegraphics[scale=0.5]{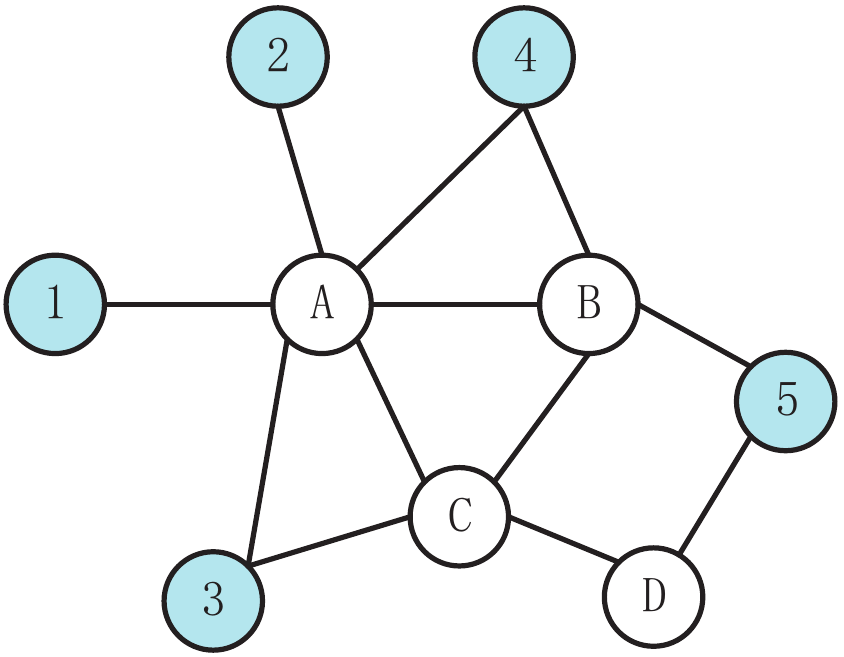}
\par\end{centering}
}\subfloat[ ]{\begin{centering}\label{fig:1-b}
\includegraphics[scale=0.5]{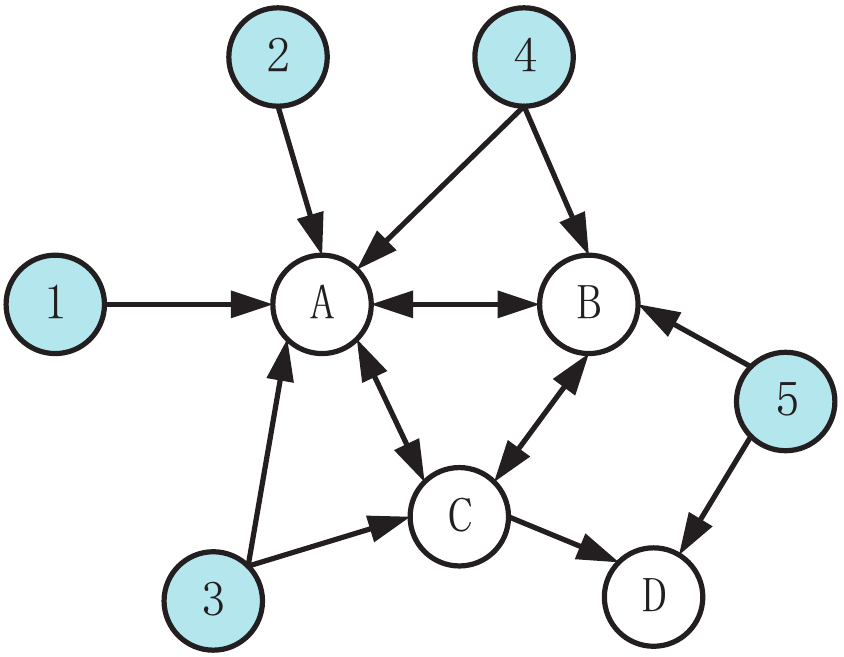}
\par\end{centering}
}
\par\end{centering}
\begin{centering}
\subfloat[ ]{\begin{centering}
\includegraphics[scale=0.5]{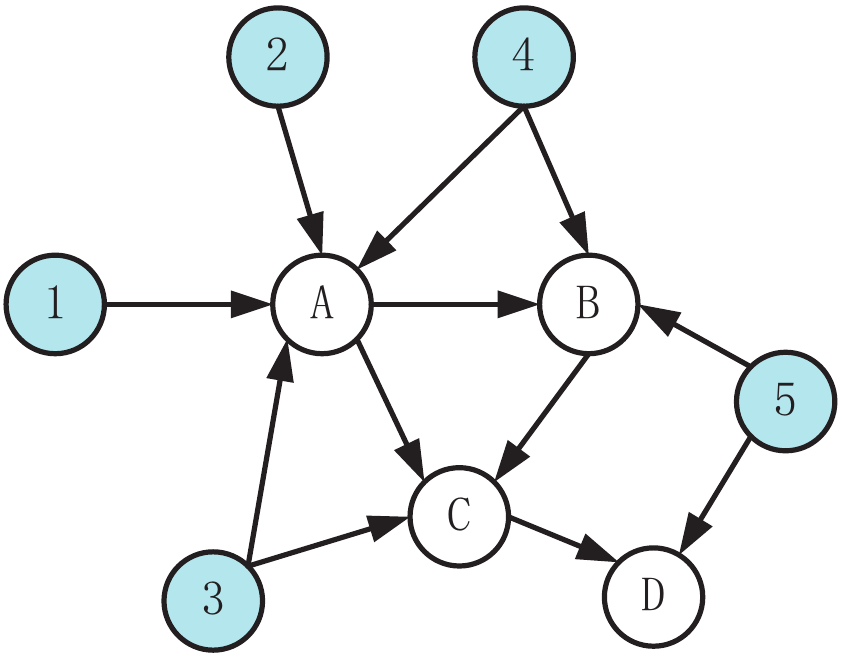}
\par\end{centering}
}
\par\end{centering}
\caption{An example of cooperative localization. where A, B, C and D are the agent nodes. while 1,
2, 3, 4 and 5 are the anchor nodes. (a) represents an undirected graph, (b) represents a bidirected graph, (c) represents an unidirectional
graph.}\label{fig:graph_models}
\end{figure}

A graphical model is a probabilistic model which employs a graph for denoting
the conditional independence structure of $N$ random variables. Typically, there are two types of graphical models, namely the directed graphical models (Bayesian networks)\cite{das2012censoring}
and the undirected graphical models (Markov networks)\cite{ihler2005nonparametric,savic2009sensor,wymeersch2009cooperative}.
As shown in Fig. \ref{fig:graph_models}, cooperative localization can be defined on either
an undirected graphical model or a directed graphical model (including unidirectional graphs and bidirected graphs). A factor
graph (FG) is capable of providing a unified interpretation of these graphical
models and lends itself to a low-complexity implementation of distributed localization.

An FG is a bipartite bidirected graph that illustrates how a complicated global function
relying on many variable factors may be simplified into the product of several
simple local functions, where each local function has fewer
variables. More specifically, an FG has a variable node representing each variable, a function
node for each local function, and an edge connecting a variable node
to a function node if and only if the variable is an argument of the local
function. The relationship between the FG and the joint distribution of all variables
can be quantified using the potential functions $\psi_{Q}$, which are
defined on each ``clique'' $Q$ of the graph. Explicitly, a clique is a subset of the complete set $V$ of variable nodes, and every two vertices in the clique are connected if the clique contains more than one vertices.
The joint PDF of all agent positions is then formulated as:
\begin{equation}
p(\mathbf{x}_{1},\ldots,\mathbf{x}_{N})\propto\underset{clique\ Q}{\prod}\psi_{Q}(\{\mathbf{x}_{q}:q\in Q\}).
\end{equation}
For the distance-based localization in our system model, the joint
\emph{a posteriori} distribution of the sensor positions can be expressed by the potential functions defined
over the set $V$ of variable nodes as follows
\begin{equation}
p(\mathbf{x}_{1},...,\mathbf{x}_{N}|\{\tilde{d}_{vi}\})\propto\underset{v\in V}{\prod}\psi_{v}(\mathbf{x}_{v})\underset{(v,i)\in E}{\prod}\psi_{vi}(\mathbf{x}_{v},\mathbf{x}_{i}),\, v\in V,
\end{equation}
where $\psi_{v}(\mathbf{x}_{v})$ represents the single-node potential,
which is defined as the  \textit{a priori} information concerning the position of sensor node $v$, i.e. we have
$\psi_{v}(\mathbf{x}_{v})=p(\mathbf{x}_{v})$. Additionally, $\psi_{vi}(\mathbf{x}_{v},\mathbf{x}_{i})$
is the pairwise potential between node $v$ and node $i$, which is
defined as $\psi_{vi}(\mathbf{x}_{v},\mathbf{x}_{i})=p_{d}(\mathbf{x}_{v},\mathbf{x}_{i})p(\tilde{d}_{vi}|\mathbf{x}_{v},\mathbf{x}_{i})$.

\begin{figure}[t]
\begin{centering}
\includegraphics[width=3.3in]{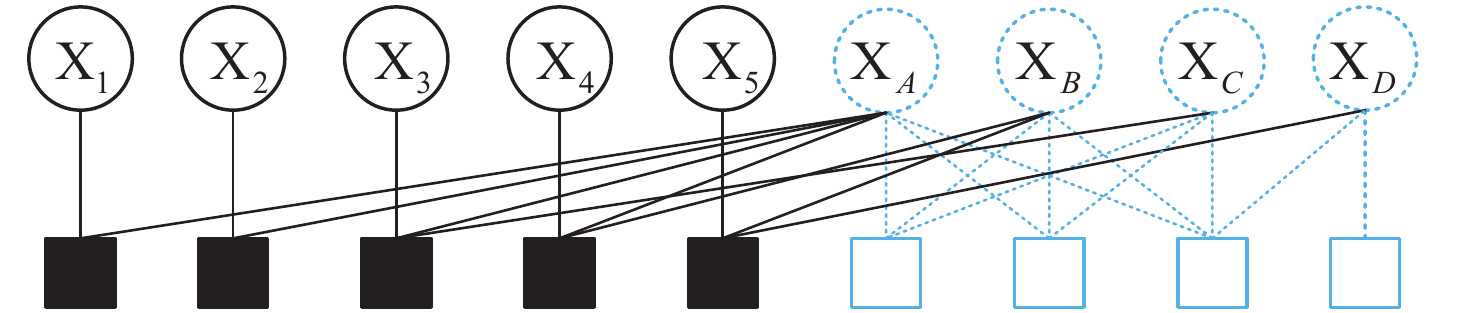}
\par\end{centering}
\caption{An FG of the example given in Fig. 1(b). Variable nodes are represented
by circles, function nodes are represented by squares. Edges that connect the agents and anchors are represented by solid lines. Edges that connect two agents are represented by dashed lines. Every two variable
nodes exchange information according to the rule specified by a function node.}
\label{fig:FG}
\end{figure}

Having defined a statistical frame work for sensor localization relying on a graph,
we can now estimate the locations of agents by applying the classic message-passing
algorithms for approximating the \textit{a posteriori} marginals of the agent positions. We consider the sensor nodes in a WSN, as exemplified in Fig. 1(b), as the variable
nodes in an FG. Correspondingly, the FG function nodes compute the message
between a pair of sensor nodes in a manner as shown in Fig. 2. The sensor nodes estimate
their beliefs (or \textit{a posteriori} marginals) by collecting all incoming
messages from the neighboring nodes during each time slot. Note that the
belief of node $i$ during time slot $t$ (i.e. in the $t$-th iteration)
is proportional to the product of the messages merging into node $i$ and to the local evidence which is determined
by the single-node potential $\psi_{i}^{t}(\mathbf{x}_{i})$. Then, the updating rule of the belief at time slot $t$ is formulated as
\begin{equation}\label{eq:update_rule}
B_{i}^{t}(\mathbf{x}_{i})\propto\psi_{i}^{t}(\mathbf{x}_{i})\underset{s\in S_{i}}{\prod}M_{si}^{t}(\mathbf{x}_{i}),
\end{equation}
where $M_{si}^{t}(\mathbf{x}_{i})$ is the message sent from node $s$
to node $i$, $s\in S_{i}$. The message $M_{si}^{t}(\mathbf{x}_{i})$ is computed
by the function nodes exemplified in Fig. \ref{fig:FG} according to the message updating rule of
\begin{eqnarray}\label{eq:message_update_rule}
M_{si}^{t}(\mathbf{x}_{i}) & = & \int_{\mathbf{x}_{s}}\psi_{s}^{t-1}(\mathbf{x}_{s})\psi_{si}^{t}(\mathbf{x}_{s},\mathbf{x}_{i})\underset{o\in S_{s}\backslash i}{\prod}M_{os}^{t-1}(\mathbf{x}_{s})\mathrm{d}\mathbf{x}_{s}  \nonumber  \\
& = & \int_{\mathbf{x}_{s}}\frac{B_{s}^{t-1}}{M_{is}^{t-1}}(\mathbf{x}_{s})\psi_{si}^{t}(\mathbf{x}_{s},\mathbf{x}_{i})\mathrm{d}\mathbf{x}_{s}.
\end{eqnarray}

The available information is then processed among the neighboring
nodes according to (\ref{eq:update_rule}) and (\ref{eq:message_update_rule}). Each variable node modifies its own positional information
by fusing the local evidence and the received messages. Provided that the belief of node $i$ converges, the variable node $i$ becomes capable of accurately estimating its position after a few
iterations. However, convergence may not be reached in those graphs which have loops.

To evaluate the uncertainty of the information propagated between the
variable nodes and function nodes, we introduce the notion of complementary entropy-ratio (CER)
for representing the deviation between two arbitrary distributions. Explicitly, the CER is
defined as
\begin{equation}
\gamma(\mathbf{x}_{i})=1-\frac{H(\mathbf{x}_{i}^{\star})}{H(\mathbf{x}_{i}|\{\mathbf{x}_{s},\tilde{d}_{si}\})},\, s\in S_{i},
\end{equation}
where $H(\mathbf{x}_{i}^{\star})$ stands for the entropy of the true distribution of the position of node
$i$, while $H(\mathbf{x}_{i}|\{\mathbf{x}_{s},\tilde{d}_{si}\})$
is the entropy of the approximate PDF of $\mathbf{x}_{i}$ estimated
by the neighboring node $s$.
Note that the location information in our approach is approximated by a nonparametric representation. As a result, it is impossible for the entropy and for the conditional entropy to become negative.
It is plausible that a lower CER represents a more accurate approximate PDF.
Before the iterative message
passing commences, the CER of anchor nodes is zero, while that of the agent nodes
is one. After several message exchanges,
a fixed point of convergence can be acquired if all messages are consistent and the
number of messages is no less than three, which implies that the principle of trilateration has to be satisfied.
Thus, we can formulate our localization inference problem
subject to additional constraints for the sake of achieving a reliable location estimation
as follows:
\begin{eqnarray}
 & \underset{\mathcal{F}_{i}}{\mathrm{min}} & \gamma(\mathbf{x}_{i})=\underset{\mathcal{F}_{i}}{\mathrm{min}}\;\;\;1-\frac{H(\mathbf{x}_{i}^{\star})}{H(\mathbf{x}_{i}|\{\mathbf{x}_{u},\tilde{d}_{ui}\})},u\in\mathcal{F}_{i},\\
 & \mathrm{s.t.} & \gamma(\mathbf{x}_{u})\leq\delta,\nonumber \\
 &  & \mathcal{F}_{i}\subset S_{i},\,|\mathcal{F}_{i}|\geq3,\nonumber
\end{eqnarray}where $\mathcal{F}_{i}$ contains the reference nodes selected for information fusion, hence it is a subset of $S_{i}$; furthermore, $\delta$ is a small value, which indicates that the
reference node employed provides reliable positional information. 
Because we have \textbf{$\gamma(\mathbf{x}_{u})\rightarrow0,\; u\in\mathcal{F}_{i},$}
$\left|\mathcal{F}_{i}\right|\geqslant3$, and $\mathcal{F}_{i}\subset S_{i}$, node $i$ has three well-positioned neighboring nodes. According
to the principle of trilateration \cite{dai2012,wymeersch2009cooperative}, the ``just-sufficiently
accurate'' position of node $i$ can be readily obtained for the next-stage process to refine it. Based on the theory
of NBP \cite{ihler2005nonparametric,sudderth2010nonparametric}, $p(\mathbf{x}_{i}|\{\mathbf{x}_{u},\tilde{d}_{ui}\})$
(i.e., the approximate PDF of node $i$ estimated by the neighboring
nodes $u\in\mathcal{F}_{i}$) closely resembles $p(\mathbf{x}_{i}^{\star})$
(i.e., the true distribution of the position of node $i$).

\subsection{Hierarchical Graphical Model}\label{sec:Hierarchical_graph_model}

In this subsection, a hierarchical graphical model considering
the quality of the positional information of the sensor nodes is presented. For
notational convenience, we further define the set of anchor indices as
$\mathcal{A}_{0}=\{j|\, j\in J\}$, and treat the set of all the anchor nodes as the root layer in the hierarchical graphical model. Additionally, both position estimation and position broadcasting are carried out by the sensor nodes. In order to control the behaviors of the sensor nodes, either the active or the inactive state is assigned to each sensor node. Note that only the active nodes are capable of estimating the positions
and broadcasting messages. In the initial step, the anchor nodes are active,
while the agent nodes remain silent. Then, when the inactive agent nodes are in connection
with the active nodes and the number of these active nodes satisfies the
threshold value, the agent nodes are gradually activated by exploiting the bootstrap
percolation strategy, which simply spreads activation from the already activated nodes to the hitherto inactive nodes. To elaborate a little further, let us define $\mathcal{A}_{l}$ as the set of agent nodes that are activated on layer $l$, where layer $l$ is defined as the specific set of agents having the same level of confidence in position
estimation. Explicitly, a coarse-grained definition of the confidence in the position estimate at node $i$ is given by
\begin{equation}\label{eq:confidence_definition}
C_f(N_{i})=\begin{cases}
1, & N_{i}=1,\\
2, & N_{i}=2,\\
3, & N_{i}\geq3,
\end{cases}
\end{equation}
where we have $N_{i}=|\mathcal{F}_{i}|$. It is readily observed from (\ref{eq:confidence_definition}) that the confidence of an agent node is closely
related to its degree of connectivity with the active nodes. Note that when
the agent node is connected to at least three active nodes,
its belief can be acquired as a steady unimodal distribution, while
a bimodal distribution is obtained when the agent node is connected
to two active nodes. If an agent node is connected to a single active node, the best possible estimates of its location constitute a ring. Therefore, using
an adaptive threshold of connection degree  is considered as the soft constraint
to divide all sensor nodes into different layers.

Let us assume that the agent nodes in the upper layers $\mathcal{A}_{m},\, m=1,\ldots,l-1$,
have been activated subject to a satisfactory connection degree constraint. Then,
all activated nodes can serve as the candidate reference nodes to facilitate the localization of the agent nodes on layer $\mathcal{A}_{l}$.
The set of candidate reference nodes given by the preceding $l$ layers is defined as
\begin{equation}\label{eq:candidate_reference_nodes_set}
\mathcal{R}^{l}=\underset{m}{\cup}\mathcal{A}_{m},\, m=0,1,\ldots,l-1.
\end{equation}
Thus, the remaining agent nodes located above the $(l-1)$th layer are included in $\mathcal{B}_{l}=I\backslash\mathcal{R}^{l}$,
and we have $\mathcal{B}_{l}\cap\mathcal{R}^{l}=\varnothing$. To assist
agent node $i\in\mathcal{B}_{l}$ in the localization, we select some neighboring nodes of the agent node $i$ from the candidate reference nodes to constitute a new set as follows:
\begin{equation}\label{eq:new_set}
\mathcal{F}_{i}^{l}=\mathcal{S}_{i}\cap\mathcal{R}^{l}.
\end{equation}
Relying on (12), $\mathcal{F}_{i}^{l}$ is regarded as the set of the activated neighboring nodes of agent node $i$ accumulated from the preceding $l$ layers. Considering the nodes' confidence in their position estimate, we
can divide the remaining agent nodes having unknown positions in the set $\mathcal{B}_{l}$ into three subsets having different connection degree
thresholds. These subsets are denoted by $\mathcal{B}_{l,c}=\{i|i\in\mathcal{B}_{l},\, C_f(|\mathcal{F}_{i}^{l}|)=c\}$, satisfying
$\mathcal{B}_{l}=\underset{c}{\cup}\mathcal{B}_{l,c},\, c=1,2,3$,
where $c$ is the connection degree threshold. The specific nonempty subset whose elements have
high confidence values is preferred for constructing $\mathcal{A}_{l}$.
Therefore, $\mathcal{A}_{l}$ can be expressed as
\begin{eqnarray}\label{eq:connection_constraint}
\mathcal{A}_{l} & = & \textrm{arg}\,\underset{c=1,2,3}{\textrm{max}}\, C_f(\mathcal{B}_{l,c}),\\
& & \text{s.t.} \;\; \mathcal{B}_{l,c}\neq\varnothing.\nonumber
\end{eqnarray}
We can see from (13) that $\mathcal{A}_{l}$ is determined by the particular subset
$\mathcal{B}_{l,c}$ which has the highest confidence in the position estimate. In the layering process, we invoke the strategy of bootstrap percolation. In contrast to the conventional bootstrap percolation which uses a fixed threshold, the proposed bootstrap percolation strategy employs an adaptive connection degree threshold.

\begin{figure}[t]
\begin{centering}
\includegraphics[width=3.5in]{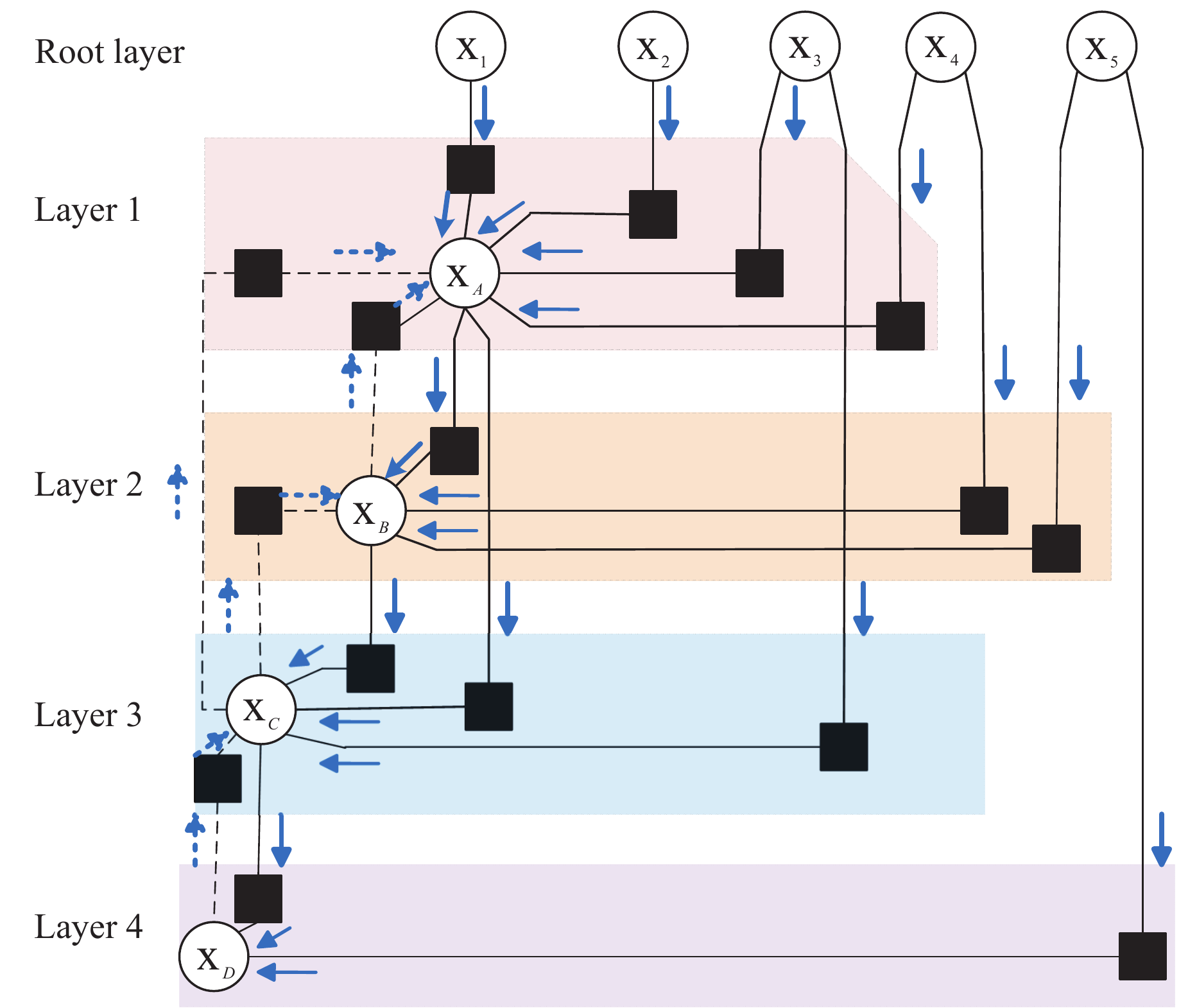}
\par\end{centering}
\caption{Information flow among the sensor nodes on the FG of Fig. 2, which characterizes the scenario shown
in Fig. 1(b). The circles stand for the variable nodes, the black squares
stand for the function nodes, the dashed arrows denote the messages originating from the lower layers, and the solid arrows represent the messages emerging from the upper layers.}
\end{figure}

According to the space-time hierarchical framework, agent nodes gradually infer
their locations. Consider the example shown in Fig. 3, where all the agent nodes $A$,
$B$, $C$ and $D$ remain silent in the initial step, while the anchor
nodes 1, 2, 3, 4 and 5 are active. The anchor nodes are collectively assigned to the root
layer, while the agent nodes $A$, $B$, $C$ and $D$ are divided into four
layers according to the connection constraints of  (\ref{eq:connection_constraint}). Agent node
$A$, which represents a class of sensor nodes connected with at least
three reference (active) nodes, is activated first. Then, agent node $A$ broadcasts
its position to the neighboring agent nodes $B$ and $C$ (note that node $D$ is not a neighboring node of $A$, as shown in Fig. 1). Afterwards, agent node $B$ is activated, since it is connected to a sufficient number of reliable
reference nodes, namely to $A$, 4 and 5. Then, agent node $B$ estimates its own position and
broadcasts the estimate to its neighboring node $C$. Subsequently,
the connection constraints of (\ref{eq:connection_constraint}) are rechecked, and node $C$ can
infer its positional distribution based on the messages sent from reference
nodes $A$, $B$ and 3. Finally, node $D$ is activated by
reference nodes $C$ and $5$. In contrast to the traditional methods of \cite{ihler2005nonparametric},
where the sensor nodes exchange their messages without control, we put emphasis
on the behavior constraints of the sensor nodes -- only the active nodes are allowed to
estimate positions and broadcast the position information. As a beneficial result, an effective scheme of managing the energy consumption and network traffic imposed by the localization process is realized. According to the hierarchical approach that relies on
the bootstrap percolation strategy described by (\ref{eq:candidate_reference_nodes_set}), (\ref{eq:new_set}), and (\ref{eq:connection_constraint}), the agent nodes satisfying
the constraint of (13) first estimate their positions, and
then assist the other hitherto inactive sensor nodes to complete localization.

In the next subsection, we will elaborate both on the information propagation rule and
on the information fusion among the sensors based on the proposed hierarchical graph model.

\subsection{The NBP Positioning Algorithm Based on the Space-Time Hierarchical-Graph}
\begin{figure}[t]
\begin{centering}
\includegraphics[width=3.5in]{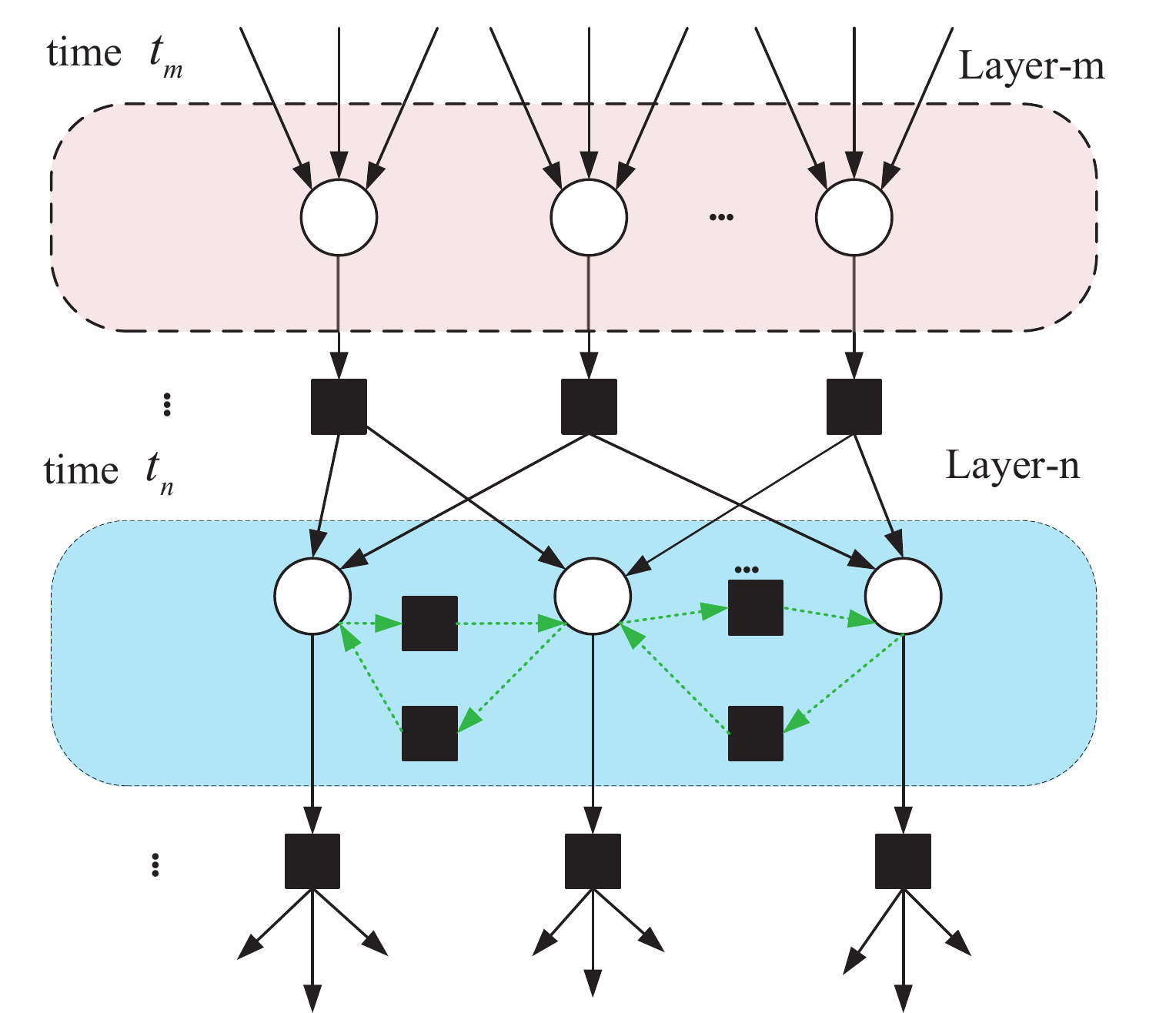}
\par\end{centering}
\centering{}\caption{The message flow on the hierarchical graph. Layer $m$ is activated at
time $t_{m}$, and layer $n$ is activated at time $t_{n}$. The solid arrows denote the inter-layer messages, and the dashed arrows represent the intra-layer messages. For inter-layer message passing,
messages always flow from upper layers to a lower layer. For the intra-layer message passing,
such as layer $n$, sensor nodes having an insufficient number of active links can exchange
messages for the sake of assisting each other in the position estimation, whereas
sensor nodes having a sufficient number of active links on layer $m$ do
not exchange messages.}
\end{figure}

In this subsection, an information propagation rule based on the hierarchical graph model of Fig. 3 is presented first. Then, an NBP algorithm is developed for the space-time hierarchical-graph. In contrast to the standard NBP positioning algorithm where all sensor nodes exchange messages iteratively, we divide the agent nodes into multiple layers and message exchange is extended from the upper layers to a certain lower layer, as shown in Section \ref{sec:Hierarchical_graph_model}. Based on such a space-time division, it can be observed that each layer is activated by its upper layers during different time slots, relying on an initialization from the root layer and on satisfying the constraint in (\ref{eq:connection_constraint}). Therefore, the proposed scheme has the potential of realizing accurate localization during different time slots. More particularly, in the proposed scheme we have three types of messages according to the corresponding layers, namely the message received from the upper layers (MUL), the message gleaned from the same layer (MSL), and those emanating from the lower layers (MLL). Based on (\ref{eq:connection_constraint}), the MUL has a high confidence, thus we should make full use of it as reliable information; the MSL serves as auxiliary information, when the agent nodes are connected to an insufficient number of reliable reference nodes; and the MLL is regarded as erroneous or misleading information. In order to take full advantage of all valuable information, whilst avoiding misleading information, we define the information propagation rule as
\begin{enumerate}
\item inter-layer message
passing: messages always propagate from upper layers to a lower layer.
\item intra-layer message passing: messages are exchanged among agent nodes, while considering the connection degree threshold $c$.
\end{enumerate}

The agent nodes connected to more than two reference nodes (i.e.
$c=3$ according to (\ref{eq:connection_constraint})) are not allowed to exchange messages (i.e. no MSL), since they are regarded
as misleading information. By contrast, the agent nodes connected to
less than three reference nodes (i.e. $c=1$ or 2 according to (\ref{eq:connection_constraint}))
are allowed to exchange their positional information for the sake of assisting each
other in improving the positioning accuracy (i.e. the MSL is required when $c=1$ or 2).

According to the proposed information propagation rule, the messages flow arranged
in a controlled manner along the above-mentioned direction is shown in Fig. 4. Based on this carefully controlled
information propagation, a \textit{spanning tree} is formulated by the
layering process. The anchor nodes constitute the root layer of the spanning
tree, and the agent nodes satisfying the constraint of (\ref{eq:connection_constraint})
constitute the leaf layers. Agent nodes at different layers are activated during different
time slots for invoking the NBP algorithm. Provided that the positional information of agents on one layer becomes convergent after several iterations, the next layer starts to execute the NBP algorithm.

\begin{figure}[t]
\begin{centering}
\includegraphics[width=3.5in]{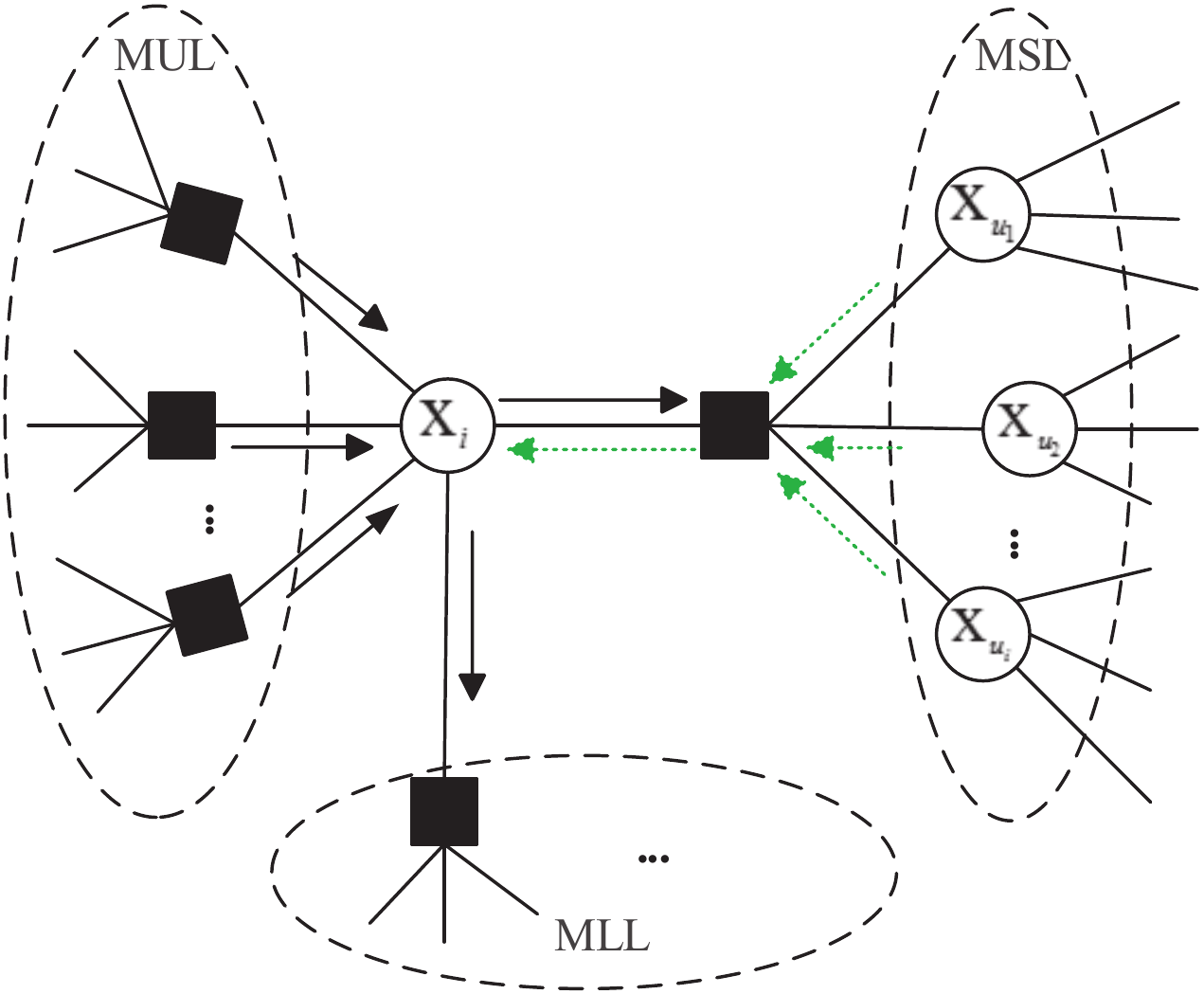}
\par\end{centering}
\centering{}\caption{The belief estimation process of an agent node. The solid arrows denote the messages from upper layers, and the dashed arrows represent the messages from the same layer. Agent node $i$ updates its belief
by fusing the incoming messages. MUL represents messages from upper
layers, MSL represents messages from the same layer, and MLL stands for
messages in lower layers. Messages are exchanged among sensor nodes
with the aid of function nodes. }
\end{figure}

Given our information propagation rule, let us now elaborate on the information
fusion rule invoked in NBP. As shown in Fig. 5, the agent node $i$ on layer
$l$ receives the MUL; when the agent node $i$ is connected with an insufficient number
(i.e. less than three) of reference nodes, it also receives the MSL.
Then, node $i$ updates its belief by fusing the messages received during time slot $t$ as follows:
\begin{equation}
b_{i}^{t}(\mathbf{x}_{i})\propto\psi_{i}^{t}(\mathbf{x}_{i})\underset{u\in\mathcal{F}_{i}}{\prod}m_{ui}^{t}(\mathbf{x}_{i}),\, i\in\mathcal{A}_{l},
\end{equation}
where the set of reference nodes is given by
\begin{equation}
\mathcal{F}_{i}=\begin{cases}
\mathcal{F}_{i}^{l}, & \,|\mathcal{F}_{i}^{l}|\geqslant3,\\
\mathcal{F}_{i}^{l}\cup A_{l}, & \,|\mathcal{F}_{i}^{l}|<3.
\end{cases}
\end{equation}
Note that $\mathcal{F}_{i}^{l}$ contains all the activated neighboring
nodes of node $i$ from the upper layers according to (\ref{eq:new_set}). When the number of reference nodes is less than three, $A_{l}$ contains the agent nodes that are in the same layer, and these nodes may assist node $i$ to update its
belief. Furthermore, according to (7), the message $m_{ui}^{t}(\mathbf{x}_{i})$
received from reference node $u$ by node $i$ at time
slot $t$ is expressed as
\begin{equation}
m_{ui}^{t}(\mathbf{x}_{i})=\int_{\mathbf{x}_{u}}\frac{b_{u}^{t-1}}{m_{iu}^{t-1}}(\mathbf{x}_{u})\psi_{ui}^{t}(\mathbf{x}_{u},\mathbf{x}_{i})\mathrm{d}\mathbf{x}_{u},\, i\in\mathcal{A}_{l},\, u\in\mathcal{F}_{i},
\end{equation}
where $b_{u}^{t-1}(\mathbf{x}_{u})$ is the belief at node $u$ in the time slot $t-1$,
representing the estimated location PDF of reference node $u$ in the $(t-1)$th time slot. In order to apply the NBP algorithm, we first obtain the nonparametric
expression of $b_{i}^{t}(\mathbf{x}_{i})$ in terms of $K$ weighted
samples as follows \cite{ihler2005nonparametric}:
\begin{equation}
\underline{b}_{i}^{t}(\mathbf{x}_{i})=\{\mathbf{x}_{i}^{(k)},w_{i}^{(k)}\},\, k=1,2,\ldots,K,
\end{equation}
which is known as the NBP marginal. Note that $\mathbf{x}_{i}^{(k)}$ is the $k$th sample of the belief $b_{i}^{t}(\mathbf{x}_{i})$, and $w_{i}^{(k)}$ is the corresponding weight of $\mathbf{x}_{i}^{(k)}$. Similarly, the nonparametric sample-based expression of $m_{ui}^{t}(\mathbf{x}_{i})$,
namely the NBP message, is given by a Gaussian mixture message generated
with the aid of $K$ weighted samples, i.e.
\begin{equation}
\underline{m}_{ui}^{t}(\mathbf{x}_{i})=\{m_{ui}^{(k)},w_{ui}^{(k)},\Lambda_{ui}\},\, u\in\mathcal{F}_{i},\, k=1,2,\ldots,K,
\end{equation}
where $m_{ui}^{(k)}$ is the $k$th sample of the message arriving from reference node $u$ to
node $i$, $w_{ui}^{(k)}$ is the corresponding weight of $m_{ui}^{(k)}$
 and $\Lambda_{ui}$ is the variance of the samples, which is approximated
as $\sigma_{ui}^{2}\mathbf{I}$.

For the sake of clarity, the detailed process of computing the NBP messages and beliefs
(marginals) is summarized in Algorithm 1 \cite{ihler2005nonparametric}.
Based on these NBP messages and beliefs, the proposed NBP-based space-time
hierarchical positioning procedure is summarized in Algorithm 2.

\begin{algorithm}[t]
\small
\caption{Computation of NBP Messages and Marginals}
\begin{algorithmic}[1]

\item $\mathbf{Compute\, NBP\, messages}$: Given $K$ weighted samples
$\{w_{v}^{(k)},\mathbf{x}_{v}^{(k)}\}$ from $b_{v}^{t}(\mathbf{x}_{v})$,
construct an approximation to $m_{si}^{t}(\mathbf{x}_{i})$ from each
activated neighbor $s\in\mathcal{S}_{i}$:

\If {$p_{d}(\mathbf{x}_{s},\mathbf{x}_{i})=1$}\;approximate $m_{si}^{t}(\mathbf{x}_{i})$ with a Gaussian mixture:

\State Draw random values for $\theta^{(k)}\backsim U[0,2\pi)$ and $n^{(k)}\sim p_{n}$, where $p_{n}$ denotes the PDF of the ranging noise.

\State Means: $m_{si}^{(k)}=\mathbf{x}_{s}^{(k)}+(\tilde{d}_{si}+n^{(k)})[\sin(\theta{}^{(k)});\,\cos(\theta{}^{(k)})]$

\State Weights: $w_{si}^{(k)}=\frac{w_{s}^{(k)}}{m_{si}^{l-1}(\mathbf{x}_{s}^{(k)})}$

\State Variance: $\Lambda_{si}=K^{-\frac{1}{3}}\cdot \text{Cov}[m_{si}^{(k)}]$
\EndIf
\item $\mathbf{Compute\, NBP\, marginals}$: Given several Gaussian mixture
messages $\underline{m}_{ui}^{t}=\{m_{ui}^{(k)},w_{ui}^{(k)},\Lambda_{ui}\}$,
$u\in\mathcal{F}_{i}$, draw samples from $b_{u}^{t-1}(\mathbf{x}_{u})$:

\For {each observed neighbor $u\in\mathcal{F}_{i}$}

\State Draw $\frac{hK}{|\mathcal{F}_{i}|}$ samples $\{\mathbf{x}_{i}^{(k)}\}$
from each message $m_{ui}^{t}$

\State Weight by $w_{i}^{(k)}=\prod_{u\in\mathcal{F}_{i}}m_{ui}^{t}(\mathbf{x}_{i}^{(k)})/\sum_{u\in\mathcal{F}_{i}}m_{ui}^{t}(\mathbf{x}_{i}^{(k)})$
\EndFor
\item From these $hK$ locations, resample $M$ times according to weights (with
replacement) to produce $K$ equal-weight samples.
\end{algorithmic}
\end{algorithm}

\begin{rem}
Algorithm 1 provides the detailed process of computing the NBP messages and
marginals at time slot $t$, where $h\geq1$ is a sampling parameter.
The parameter $h$ is adjustable. A lager $h$ indicates relying on more fused information and on increased processing time.
\end{rem}

\begin{algorithm}[t]
\small
\caption{The Space-Time Hierarchical Positioning Algorithm}
\begin{algorithmic}[1]
\item \textbf{Initialization}: The root layer is $\mathcal{A}_{0}=J$,
and the reference nodes set is $\mathcal{R}^{0}=J$.

\For {$l=1$ to $L$ \textbf}\;\{iteration index\}

\State $\mathbf{Determine\, sensors\, to\, be\, located\, on\, layer\,}l$ : Select
the agent nodes in $\mathcal{A}_{l}$ using (\ref{eq:connection_constraint}).

\For {$i\in\mathcal{A}_{l}$}\; \{the reference nodes set of node $i$ is the set $\mathcal{F}_{i}=\mathcal{F}_{i}^{l}=\mathcal{S}_{i}\cap\mathcal{R}^{l}$\}

\If {$C_f(\mathcal{A}_{l})=3$}

\State Intra-layer messages are not in use

\ElsIf {$C_f(\mathcal{A}_{l})\leq2$}

\State Add assistant reference nodes from the intra-layer nodes using $\mathcal{F}_{i}\leftarrow\mathcal{F}_{i}^{l}\cup\mathcal{A}_{l}$
\EndIf
\EndFor
\EndFor

\item $\mathbf{Information\, fusion}$: Sensors in $\mathcal{A}_{l}$
receive messages from the preceding $l$ layers $\mathcal{R}^{l}$, and update their beliefs.

\For {$i\in\mathcal{A}_{l}$}\;\{start the NBP computation using Algorithm 1 in parallel\}
\For {$t=1:T$}
\State Compute NBP messages from its reference nodes $\mathcal{F}_{i}$
\State Compute NBP marginals for agent $i$
\EndFor
\EndFor
\item $\mathbf{Reference\, nodes\, updating}$: Update the reference
nodes set using $\mathcal{R}^{l+1}\leftarrow\mathcal{R}^{l}\cup\mathcal{A}_{l}$

\end{algorithmic}
\end{algorithm}

\begin{rem}
By exploiting a heuristic search relying on the bootstrap percolation strategy,
the agent nodes are divided into multiple layers. Thus, a space division of the positions is obtained,
which is capable of mitigating the influence of loops in a graph. Based on the proposed
information propagation rule, the messages flow in controlled directions
and the MLL is regarded as misleading information, which is censored. Additionally,
localization estimation is implemented in a layer-by-layer manner, where the agents
gradually infer their positions by selecting reference nodes from the
upper layers. In this way, the network traffic overhead can be dramatically reduced. Moreover, for the agent nodes that have an insufficient number of reliable reference nodes, the messages can be exchanged among agents on the same layer (intra-layer message passing). As a result, all helpful information can be exploited for improving the localization accuracy.
\end{rem}

\section{Computational Complexity and Network Traffic Overhead}

In this section, we analyze both the  computational complexity and network traffic overhead of the proposed NBP-based
space-time hierarchical positioning algorithm. Both the computational complexity and the network
traffic are related to the number of actively participating links on the graph,
hence we adopt the average number of links used in the entire network as an indirect metric for comparing the complexity of different NBP algorithms. In particular, a comparison is made among the proposed algorithm, the standard NBP algorithm and the low-complexity
NBP defined over spanning trees with breadth first search (NBP-BFS)\cite{vladimir2010indoor}.
We consider $M$ anchor nodes placed at fixed positions and $N$ agent nodes
randomly distributed over an $a$$\times$$a$ square area. The communication
range is given by $R=pa,$ where $p$ is a proportionality coefficient.
Then $\bar{N}_{\mathrm{msg}}$, which is the average number of messages fused by an agent node, can be written as
\begin{equation}
\bar{N}_{\mathrm{msg}}=\frac{N}{a^{2}}\times\pi(pa)^{2}=N\pi p^{2}.
\end{equation}
 In the standard NBP algorithm, all messages are employed for information
fusion. Thus, the computational complexity of the standard NBP algorithm
is characterized by
\begin{equation}
O_{\mathrm{NBP}}=N\cdot\bar{N}_{\mathrm{msg}}=N^{2}\pi p^{2}.
\end{equation}
When using the NBP-BFS algorithm, $N-1$ edges are retained in the graph
and there are two messages on each edge, thus the computational complexity
of the NBP-BFS is determined by
\begin{equation}
O_{\mathrm{BFS}}=2(N-1).
\end{equation}
\begin{table}[t]\arrayrulecolor{black}
\caption{Complexity Comparison in Terms of the Average Number of Links Used in the Entire Network}\label{table:complexity_compare}
\centering{}%
\begin{tabular}{c|ccc}
\hline
Algorithm & NBP & NBP-BFS & Hierarchical NBP\tabularnewline
\hline
Complexity & $\pi N^{2}p^{2}$ & $2N-2$ & $\frac{\pi}{L^{2}}N^{2}p^{2}$\tabularnewline
\hline
\end{tabular}
\end{table}

By comparison, as far as the proposed hierarchical NBP scheme is concerned, all the
sensor nodes are divided into $L$ layers. Therefore, the computational
complexity of the proposed algorithm in each iteration is approximately determined by
\begin{equation}
O_{\mathrm{layer}}=\bar{N}\cdot\bar{N}_{\mathrm{layer}}=\frac{N}{L}\cdot\frac{N\pi p^{2}}{L}=\frac{1}{L^{2}}\cdot N^{2}\pi p^{2}=\frac{O_{\mathrm{NBP}}}{L^{2}},
\end{equation}
where $\bar{N}$ stands for the average number of agent nodes in
each layer, $\bar{N}_{\mathrm{layer}}$ represents the average number
of messages that have to be fused. In particular, if $L=1$,
the proposed algorithm shares the same complexity as the standard
NBP. The complexity comparison results of the three algorithms are summarized
in Table \ref{table:complexity_compare}.

\begin{figure}[t]
\begin{centering}
\includegraphics[width=3.5in]{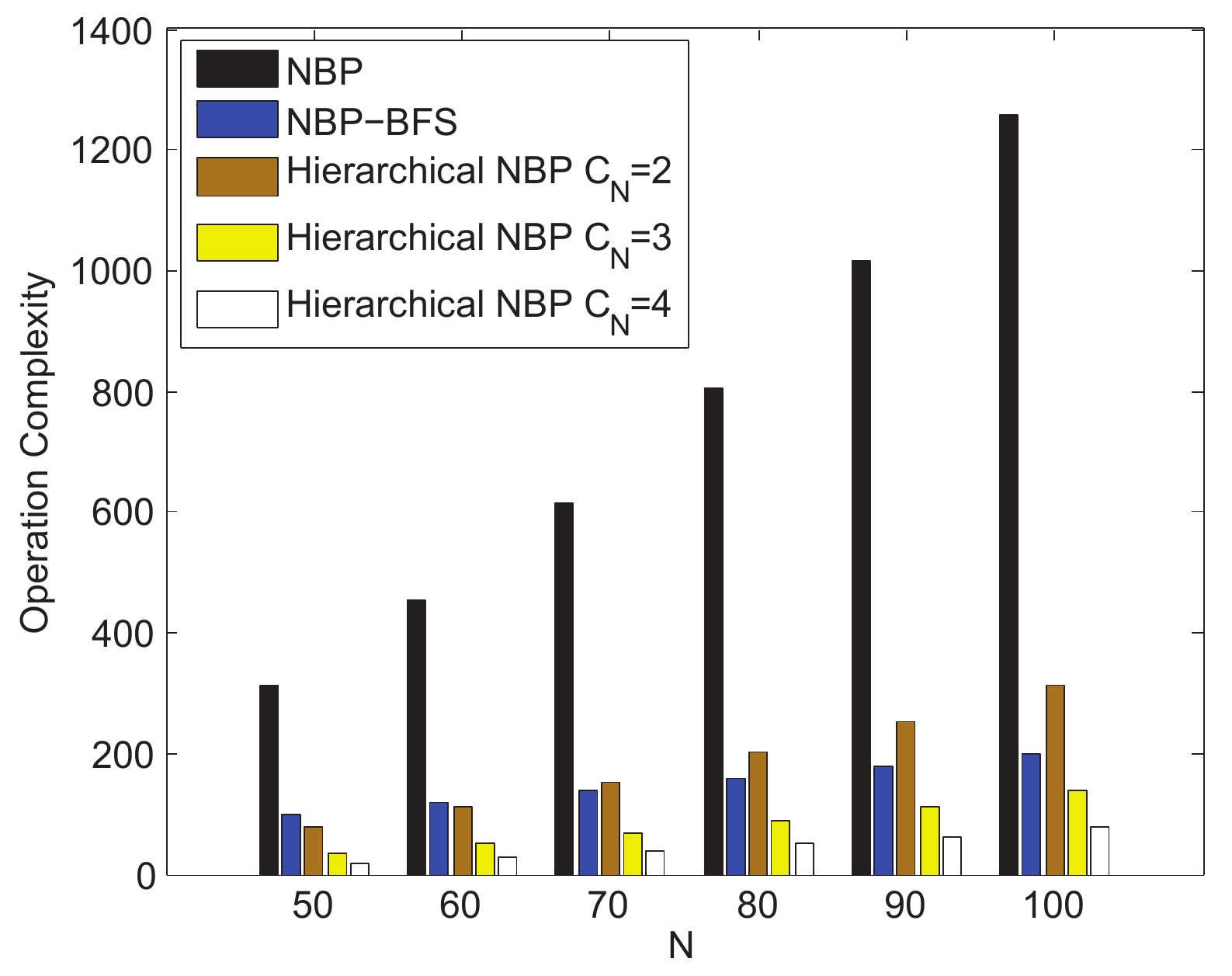}
\par\end{centering}
\caption{Complexity comparison of different NBP algorithms in terms of the average number of links used in the whole network, where $p=0.2$ is fixed and $N$ changes from 50 to 100.}
\end{figure}
\begin{figure}[t]
\begin{centering}
\includegraphics[width=3.5in]{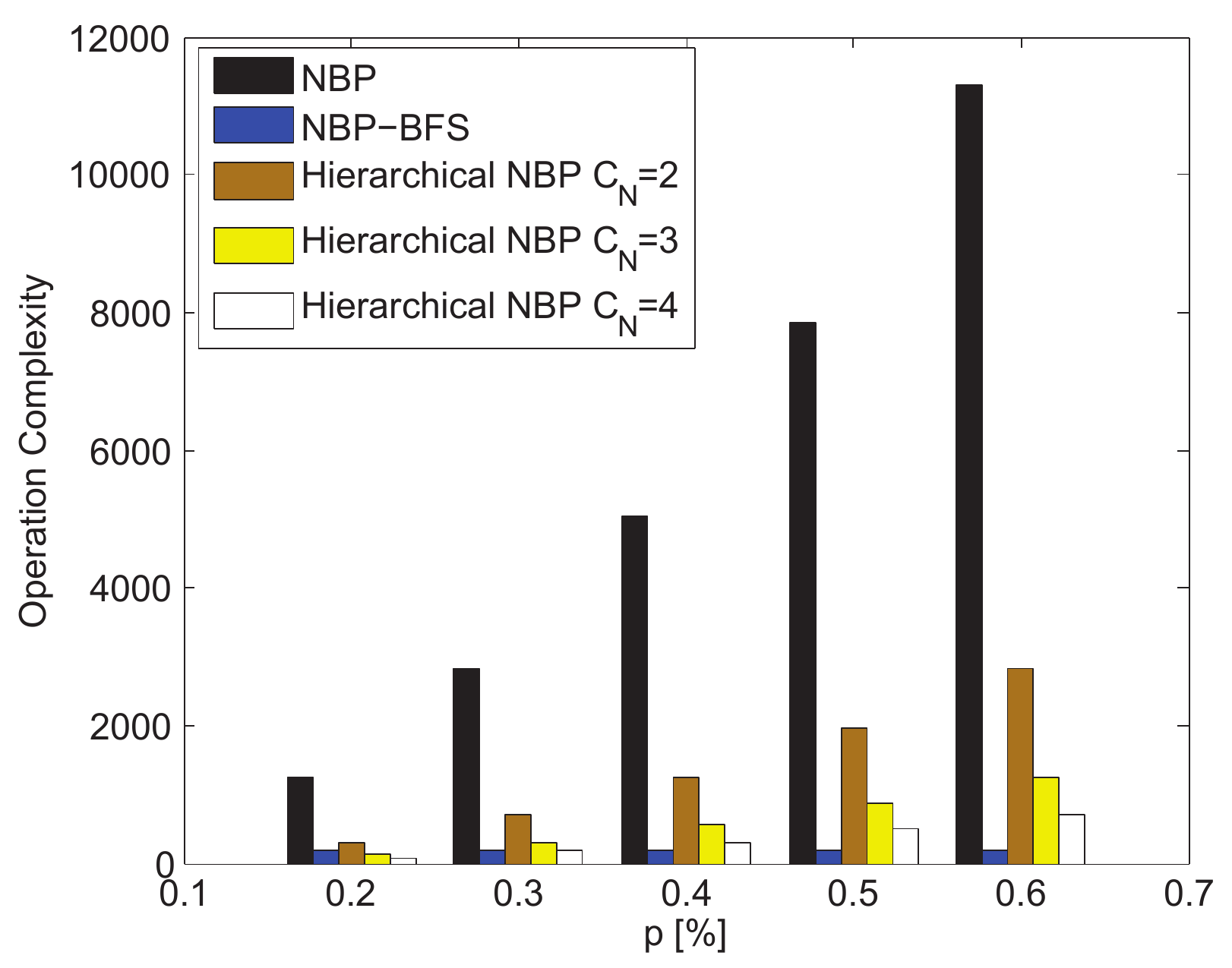}
\par\end{centering}
\caption{Complexity comparison of different NBP algorithms in terms of the average number of links used in the whole network, where $N=100$ is fixed and $p$ changes from 0.2 to 0.6.}
\end{figure}

If we have $O_{\mathrm{layer}}<O_{\mathrm{BFS}}$, the following conditions
should be satisfied,
\[
\frac{N^{2}\pi p^{2}}{L^{2}}<2(N-1),\,\,\,\frac{N\pi p^{2}}{L^{2}}<2,\,\,\textrm{and}\,\, N\pi p^{2}<2L^{2},
\]
which implies that the proposed algorithm tends to exhibit a reduced complexity
when more layers are created. As shown in Fig. 6, we consider a
typical large-scale network\cite{wymeersch2009cooperative} having 13
anchors distributed in a square area of $100$m$\times$$100$m, i.e. $a =100$, and
the transmission radius is fixed to $R=20$m, hence $p =0.2$ and $N$ changes from 50 to 100. It can be observed from Fig. 6 that when the number of agent nodes increases, the complexity of the standard NBP also increases, since there are
more messages that have to be fused, while the proposed algorithm
dramatically reduces the complexity. Fig. 7 shows our complexity
comparison of the different NBP algorithms subject to the fixed $N=100$ and to the transmission radius varying from 20m to 60m, i.e. $p$ ranges from $0.2$ to $0.6$. We can see that upon increasing the transmission radius, the average number of links
in the network increases, hence resulting in an increased complexity for the NBP.
By comparison, the proposed hierarchical scheme significantly reduces
the number of participating links, thus it has a reduced complexity. Additionally, we can see from Fig. 7 that when there are less layers, the proposed algorithm suffers from a higher complexity than NBP-BFS. Moreover, when the number of layers $L$
increases, the complexity of the proposed algorithm decreases and it has the potential to become lower than that of NBP-BFS.

\section{Simulation Results and Discussions}

In this section, numerical simulations are carried out for evaluating the performance
of the different positioning algorithms considered. In summary, compared to the standard
NBP algorithm, the proposed space-time hierarchical localization scheme
attains a performance gain by censoring the MLL that contains misleading information.
In contrast to the low-complexity NBP-BFS of \cite{vladimir2010indoor}, which runs the NBP on a spanning tree without considering the appropriate number of reference nodes, the proposed scheme runs NBP on a space-time hierarchical-graph, which ensures that most of the agent nodes share a sufficiently high number of reference nodes for localization, thus a performance gain is achieved.

\subsection{Simulation Setup}
\begin{figure}[t]
\centering{}\includegraphics[width=3.5in]{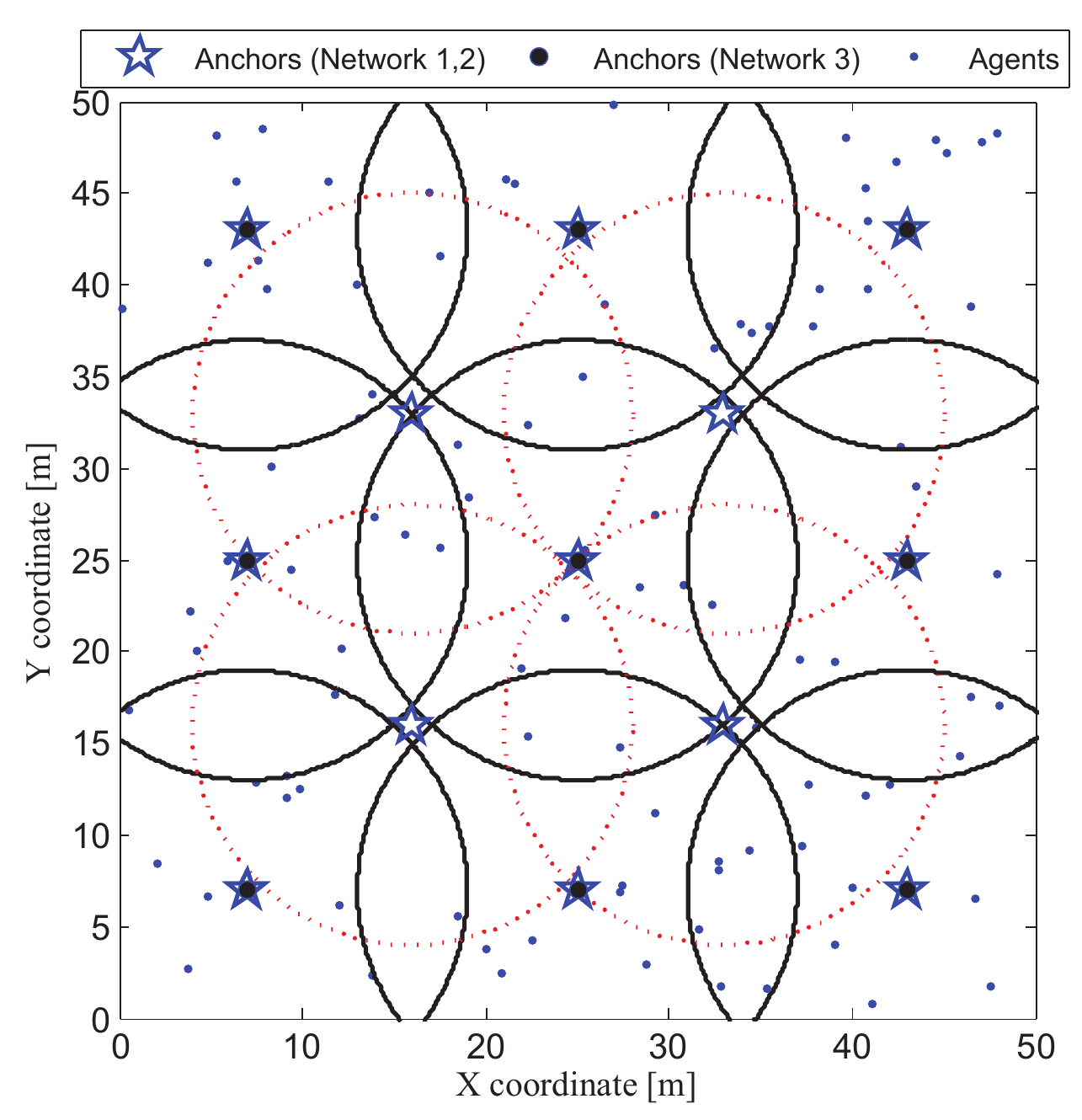}\caption{\label{deployment}Deployment of anchor nodes. The circles indicate the radio range of the anchors having a transmission radius of $R=12$m.}
\end{figure}
\begin{table}[t]\arrayrulecolor{black}
\centering{}\caption{Deployments of different networks}
\begin{tabular}{cccc}
\hline
 & Network 1 & Network 2 & Network 3\tabularnewline
\hline
Agents & 13 & 13 & 9\tabularnewline
Anchors & 100 & 50 & 100\tabularnewline
\hline
\end{tabular}
\end{table}

We consider a $50$m$\times$$50$m square area in our simulations, where
a number of anchor nodes are placed at fixed positions and a large number of agent nodes
are randomly distributed across the area. The network deployment parameters are summarized in Table III, and the \textit{a priori} positions of the anchors are shown in Fig. \ref{deployment}, where
'{\footnotesize \FiveStarOpen{}}' denotes the locations of the anchor
nodes in Network 1 and Network 2; '$\bullet$' denotes the locations
of the anchors in Network 3; and '$\cdot$' denotes the locations
of the agents.
The range measurement noise is modeled using the Gaussian distribution with zero mean and a standard deviation linearly depending on the distance. Thus the estimated distance is given by
$\tilde{d}=d+n_{s}$, $n_{s}\sim\mathcal{N}(0,(\sigma_{0}+k_{\sigma}\times d)^{2})$,
where $\sigma_{0}$ and $k_{\sigma}$ are constant values. In the simulations, it is assumed that $\sigma_{0}=0.2$m and $k_{\sigma}=0.01$. The transmission radius is $12$m, and we use $K=200$ samples for approximating the beliefs and messages. The central processing unit (CPU) utilized for the simulations is an Intel Core i5-4570 CPU. Two criteria are utilized
for evaluating the performance, including the CPU running time\footnote{Apart from the metric of ``the average number of links used in the entire network'', which is used in Table II, the CPU running time is invoked in the simulations as a complementary metric for numerically characterizing the computational complexity.} and the error distribution CDF, which is given by
\begin{equation}
\mathsf{CDF}(e_{th})=\mathbb{E}\{\mathbb{I}\{||\mathbf{x}-\hat{\mathbf{x}}||<e_{th}\}\},
\end{equation}
where $\mathbf{x}=[\mathbf{x}_{1},\ldots,\mathbf{x}_{N}]$, $\hat{\mathbf{x}}=[\hat{\mathbf{x}}_{1},\ldots,\hat{\mathbf{x}}_{N}]$, and $e_{th}$ is the positioning error threshold of an agent. In the simulations, the estimated parameters are exchanged
among the sensor nodes 10 times. All the positioning errors are obtained for 1000 randomly generated networks.

\subsection{Performance Results and Discussions}

Fig. 9, Fig. 10 and Fig. 11 show the positioning performance of the proposed NBP based space-time hierarchical positioning algorithm subject to differently initialized connection degree thresholds given in (\ref{eq:connection_constraint}) for the three networks specified in Table III, respectively.
The positioning performance of Network 1 is better than that of the other two networks, because Network 1 has enough anchor nodes and dense agents.
Note that when the connection degree threshold is initialized to zero, the proposed space-time hierarchical NBP reduces to the standard NBP.
It can be seen that the proposed space-time hierarchical NBP algorithm using a connection degree threshold higher than two outperforms the standard
NBP algorithm (i.e. the connection degree threshold is initialized as $c=0$), due to its resilience against error propagation. Additionally, we evaluated the scenario where different hierarchical graphs are generated corresponding to differently initialized connection degree thresholds. We observe from Fig. 9, Fig. 10 and Fig. 11 that there are no significant positioning performance differences for the proposed space-time hierarchical NBP algorithm, when the connection degree thresholds are increased from zero to two. Moreover, the space-time hierarchical NBP algorithm exhibits a significantly improved performance when the connection degree threshold is higher than two.
Fig. 12 shows that the CPU running time of the proposed space-time hierarchical NBP algorithm can
be reduced when more layers are created, which is consistent with the complexity analysis results shown in Section IV.
\begin{figure}[t]
\begin{centering}
\includegraphics[width=3.5in]{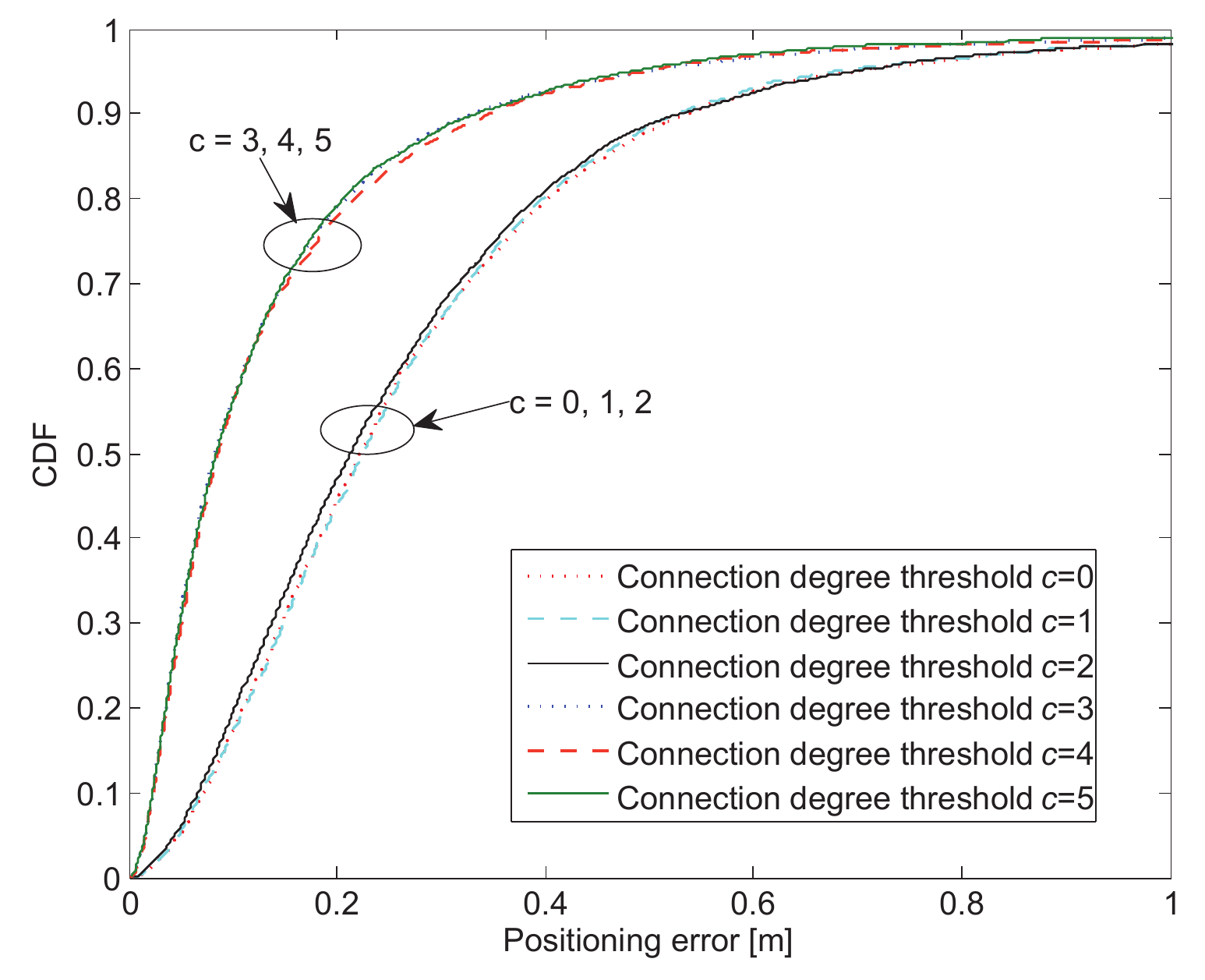}
\caption{Comparison of positioning error CDFs of hierarchical NBP under differently initialized connection thresholds for Network 1 specified in Table III.}
\par\end{centering}
\end{figure}

\begin{figure}[t]
\begin{centering}
\includegraphics[width=3.5in]{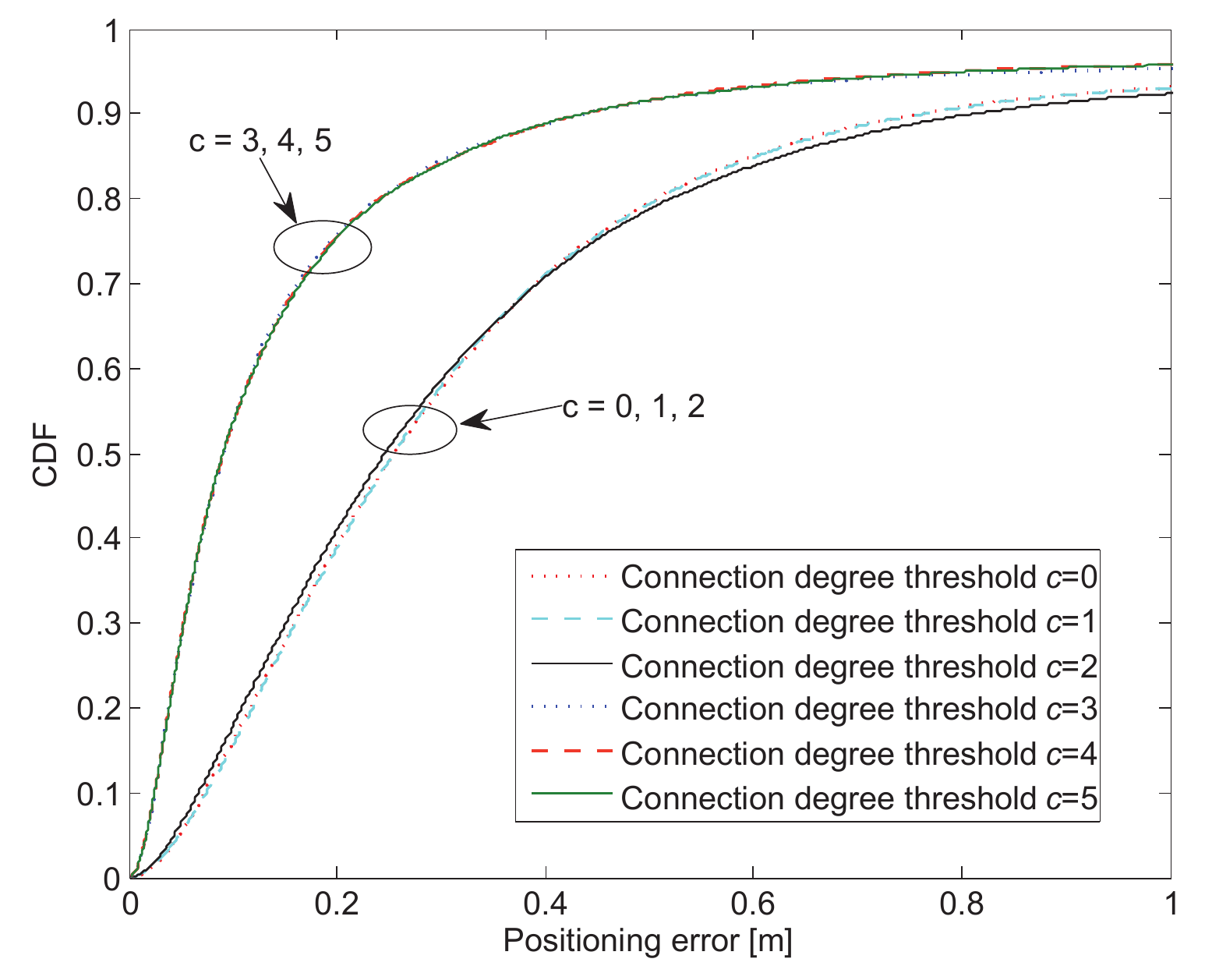}
\caption{Comparison of positioning error CDFs of hierarchical NBP under differently initialized connection thresholds for Network 2 specified in Table III.}
\par\end{centering}
\end{figure}

\begin{figure}[t]
\centering{}
\includegraphics[width=3.5in]{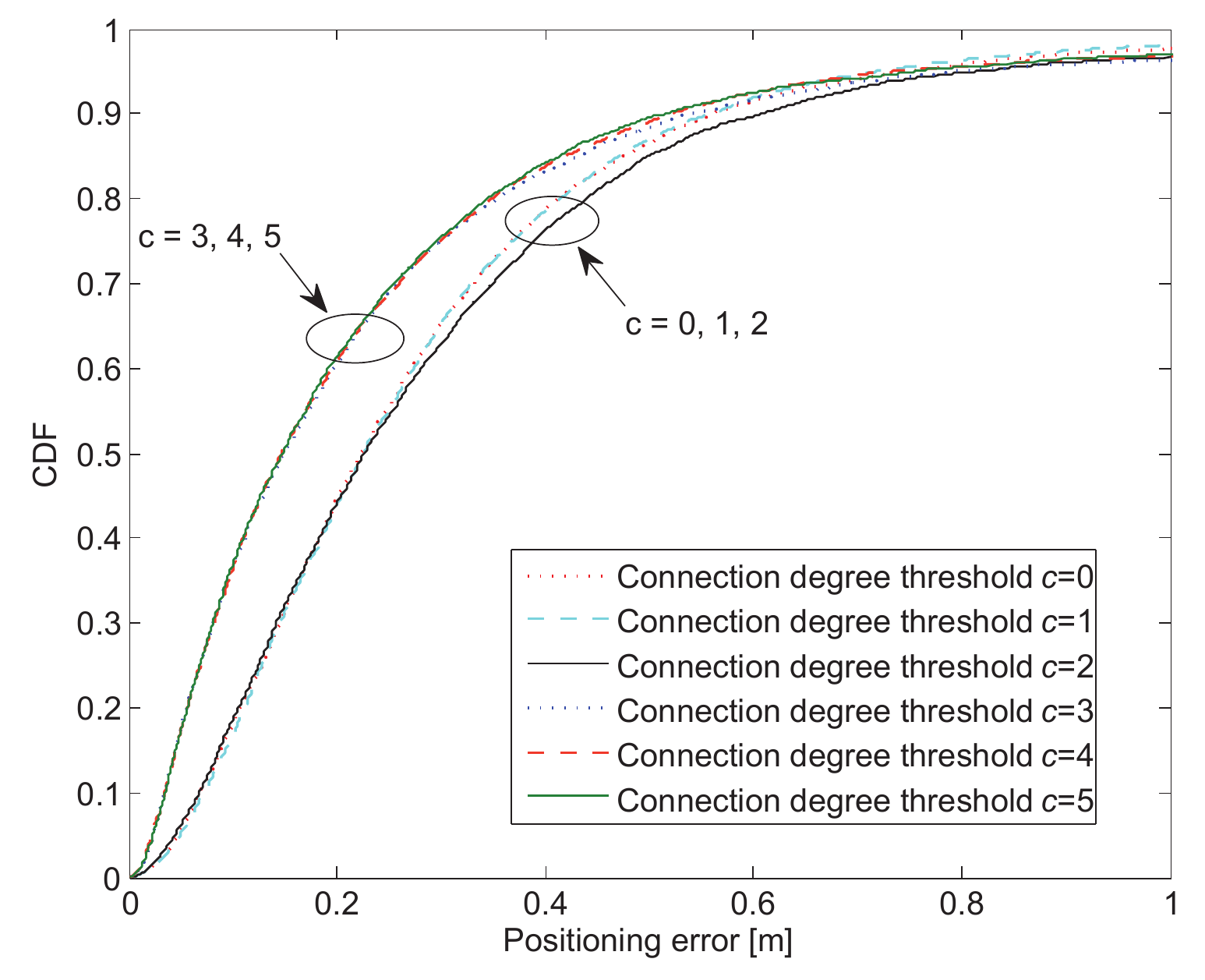}
\caption{Comparison of positioning error CDFs of hierarchical NBP under differently initialized connection thresholds for Network 3 of Table III.}
\end{figure}
\begin{figure}[tbh]
\begin{centering}
\includegraphics[width=3.5in]{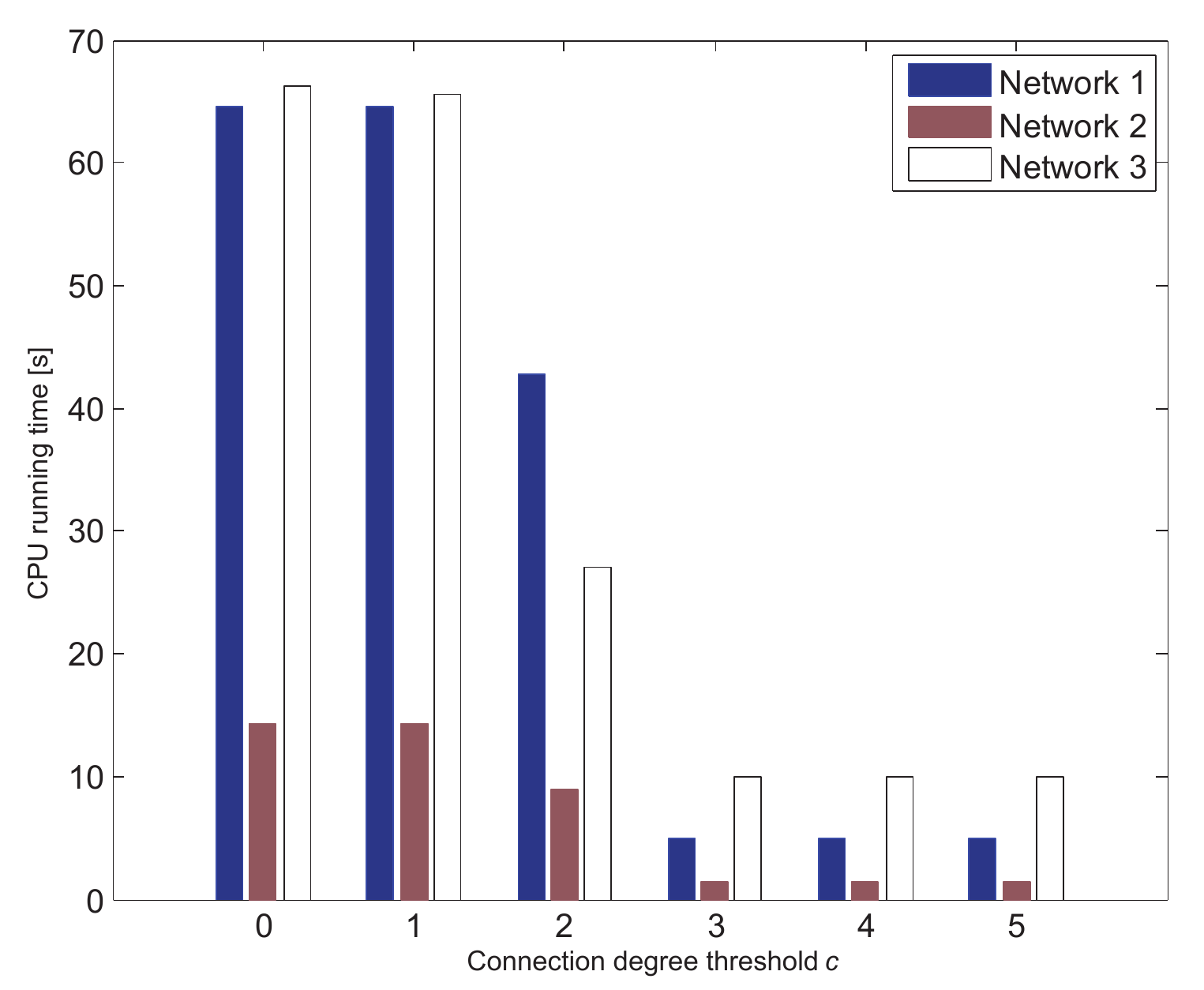}
\caption{CPU running time comparison of hierarchical NBP under different connection thresholds for the three networks of Table III, respectively.}
\par\end{centering}
\end{figure}

\begin{figure}[t]
\begin{centering}
\includegraphics[width=3.5in]{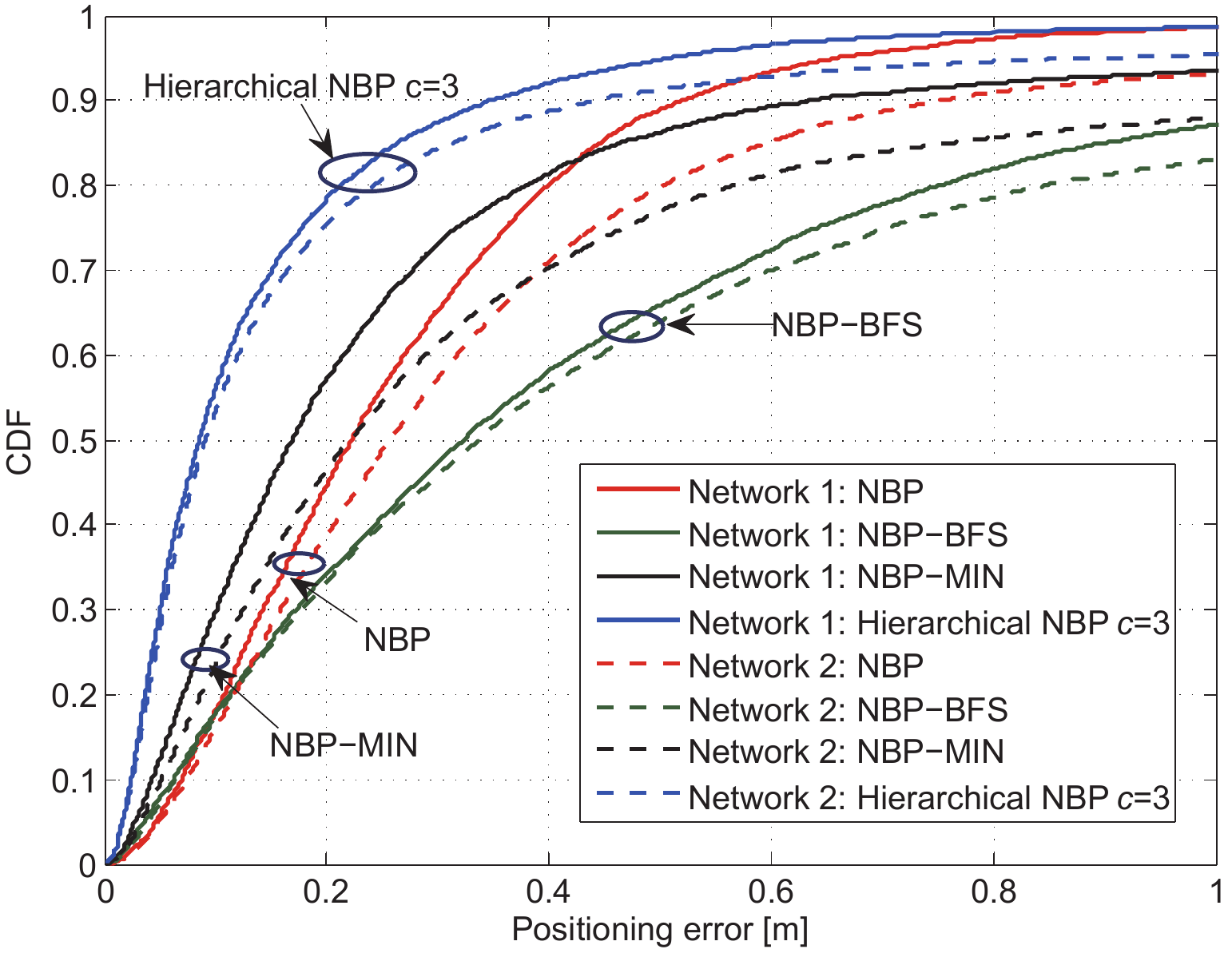}
\caption{CDF comparison of different NBP algorithms for Network 1 and Network 2 of Table III.}
\par\end{centering}
\end{figure}

The CDF comparison of different NBP algorithms for Network 1 and Network 2 is shown in Fig. 13, where it is observed that the CDF of the NBP-BFS \cite{vladimir2010indoor} indicates a
poor positioning performance, because insufficient information is gleaned.
By comparison, the space-time hierarchical NBP (where the connection degree threshold is initialized as three) makes full use of all beneficial information and mitigates error propagation, thus attaining an improved positioning error performance. We can see that the proposed space-time hierarchical NBP shows the best performance, and most of the sensor nodes are located within a positioning error of  less than $1$m.
In addition to its beneficial performance gain, we can see from Table IV that the proposed space-time hierarchical NBP algorithm significantly reduces the complexity, hence incurring a reduced CPU running time. We observe that with the aid of the proposed space-time hierarchical strategy, the NBP can be executed roughly 15 and 10 times faster than the standard NBP for Network 1 and Network 2, respectively.

\begin{table*}[t]
\small
\centering{}\caption{CPU running time comparison of different positioning algorithms for
Network 1 and Network 2.}
\begin{tabular}{ccccc}
\hline
Positioning algorithms & NBP\cite{sudderth2010nonparametric,ihler2005nonparametric} & NBP-BFS\cite{vladimir2010indoor} & NBP-MIN\cite{Xiaopeng_2015:NBP_MIN} & Hierarchical NBP\tabularnewline
\hline
CPU running time of Network 1 {[}s{]} & 71.03  & 10.53  & 15.24  & 4.80\tabularnewline
\hline
CPU running time of Network 2 {[}s{]} & 15.01 & 4.59 & 5.94 & 1.47\tabularnewline
\hline
\end{tabular}
\end{table*}

\section{Conclusions}

In this paper, a space-time hierarchical-graph was proposed for efficient
cooperative positioning in WSNs. Specifically, a bootstrap percolation strategy relying on
soft constraints was utilized for controlling the actions of sensor
nodes. As a benefit of the hierarchical mechanism, agent nodes that satisfy our
connection constraints are activated gradually for estimating their
positions, while the sensors that do not satisfy the connection constraints
remain silent. Hence, the energy consumption/network traffic overhead imposed by the localization process is significantly reduced compared to that of the benchmarking schemes considered. Additionally, an information propagation rule was designed by exploiting the hierarchical
graph. As a result, the messages always flow in the direction from the sensor
nodes of higher confidence to those having a lower confidence. Therefore, the
error propagation can be effectively mitigated, yielding an accelerated
positioning convergence. In our future work, the bootstrap percolation strategy relying on soft constraints and other features of WSNs (e.g. multi-hop)  may be exploited in network forming, which is expected to provide an improved performance gain or to reduce the  network traffic overhead imposed by the localization process.

\bibliographystyle{IEEEtran}
\bibliography{reference}

\begin{IEEEbiography}[{\includegraphics[width=1in,height=1.25in,clip,keepaspectratio]{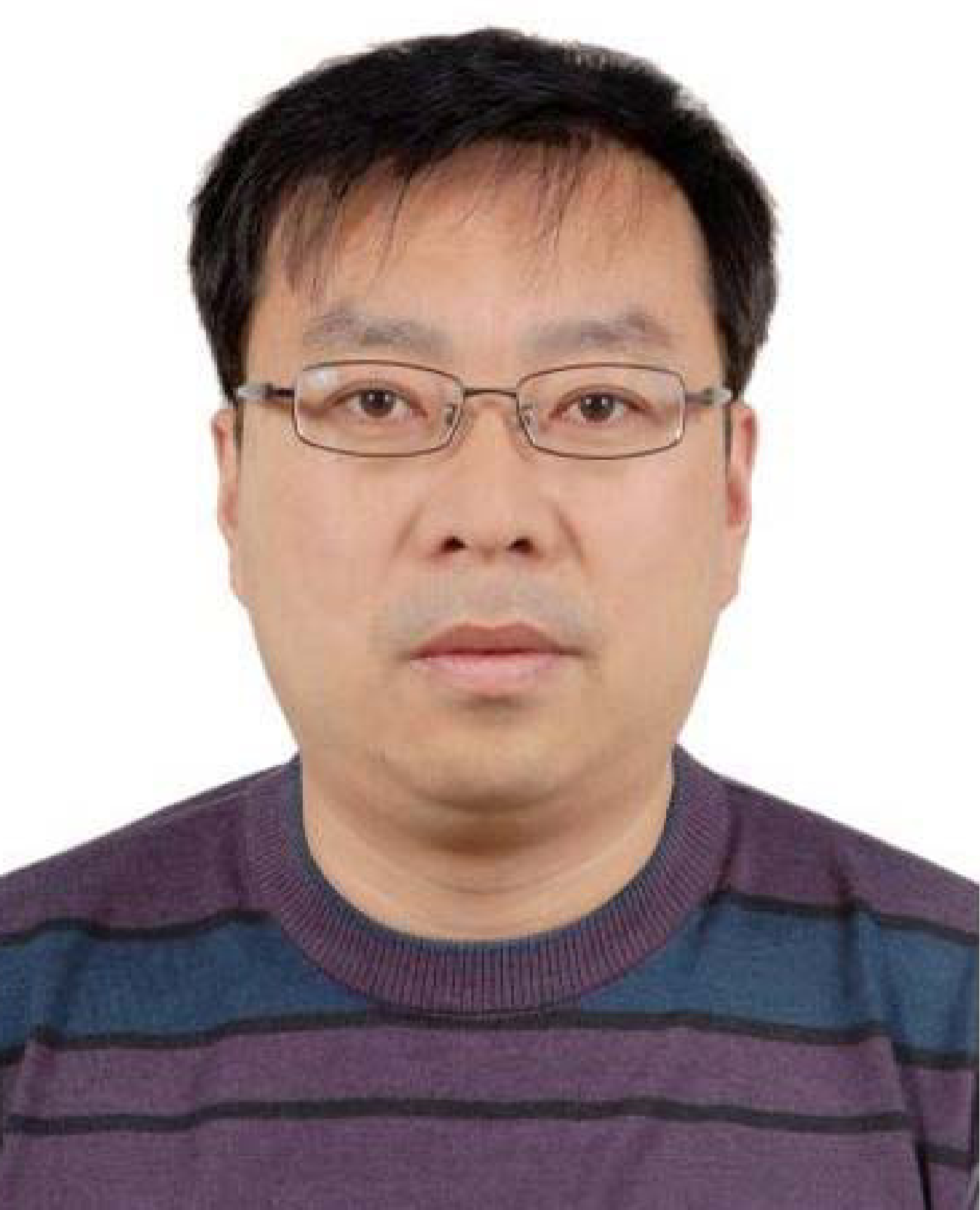}}] {Tiejun Lv}
(M'08-SM'12) received the M.S. and
Ph.D. degrees in electronic engineering from the
University of Electronic Science and Technology
of China (UESTC), Chengdu, China, in 1997 and
2000, respectively. From January 2001 to January
2003, he was a Postdoctoral Fellow with Tsinghua
University, Beijing, China. In 2005, he became a Full
Professor with the School of Information and Communication
Engineering, Beijing University of Posts and Telecommunications
(BUPT). From September 2008 to March 2009, he was a Visiting Professor with
the Department of Electrical Engineering, Stanford University, Stanford, CA. 
He is the author of more than 200 published technical papers on
the physical layer of wireless mobile communications. His current research
interests include signal processing, communications theory and networking.
Dr. Lv is also a Senior Member of the Chinese Electronics Association.
He was the recipient of the Program for New Century Excellent Talents in
University Award from the Ministry of Education, China, in 2006.
\end{IEEEbiography}

\begin{IEEEbiography}[{\includegraphics[width=1in,height=1.25in,clip,keepaspectratio]{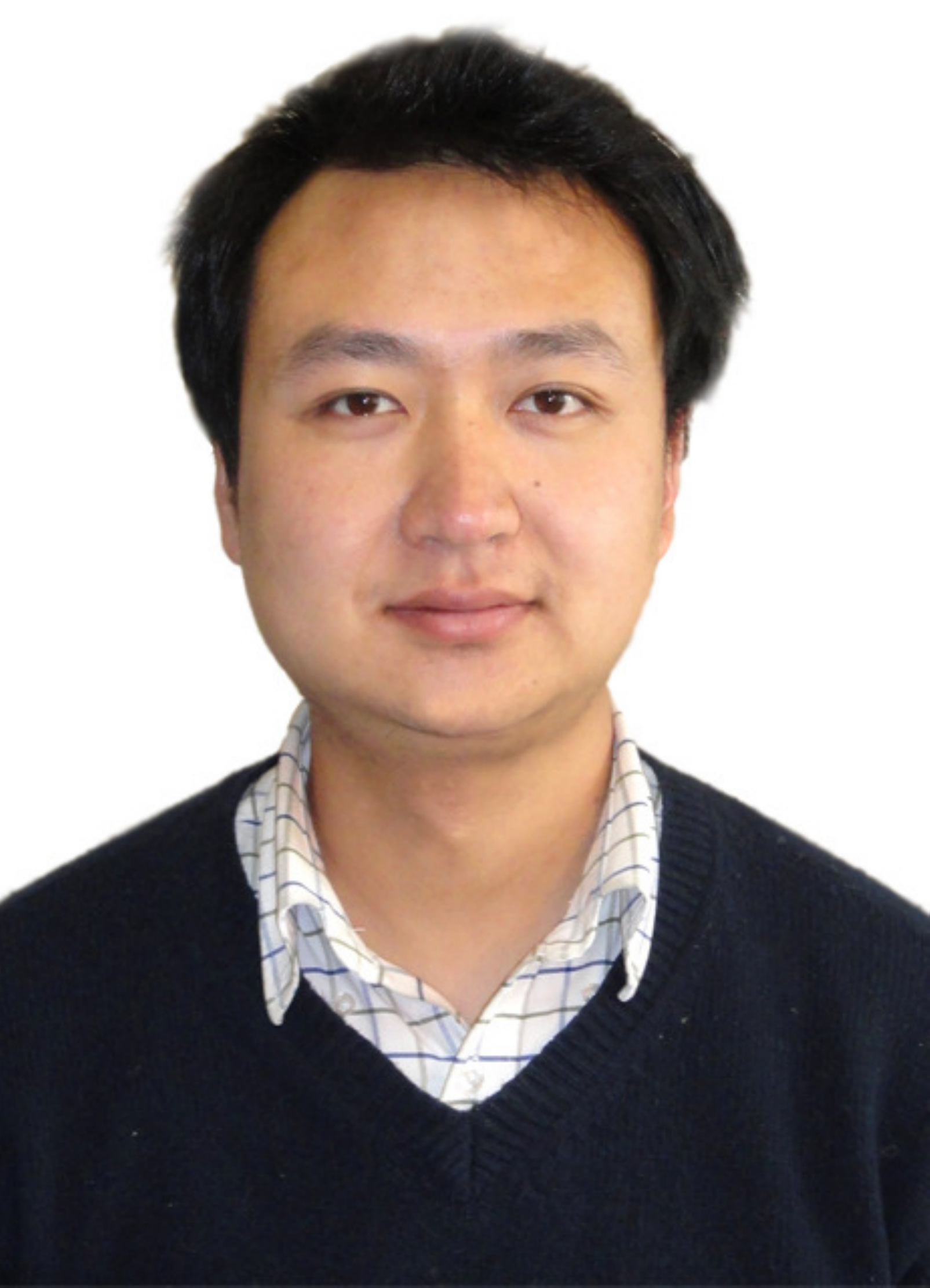}}] {Hui Gao}
(S'10-M'13) received the B. Eng. degree in information engineering and the Ph.D. degree in signal and information processing from Beijing University of Posts and Telecommunications (BUPT), Beijing, China, in July 2007 and July 2012, respectively. From May 2009 to June 2012, he was a Research Assistant at the Wireless and Mobile Communications Technology R$\&$D Center, Tsinghua University, Beijing. From April 2012 to June 2012, he was also a Research Assistant with Singapore University of Technology and Design, Singapore, where he later worked as a Postdoctoral Researcher from July 2012 to February 2014. He is currently an Assistant Professor with the School of Information and Communication Engineering, BUPT. His research interests include massive multiple-input-multiple-output systems, cooperative communications, and ultrawideband wireless communications.
\end{IEEEbiography}

\begin{IEEEbiography}[{\includegraphics[width=1in,height=1.25in,clip,keepaspectratio]{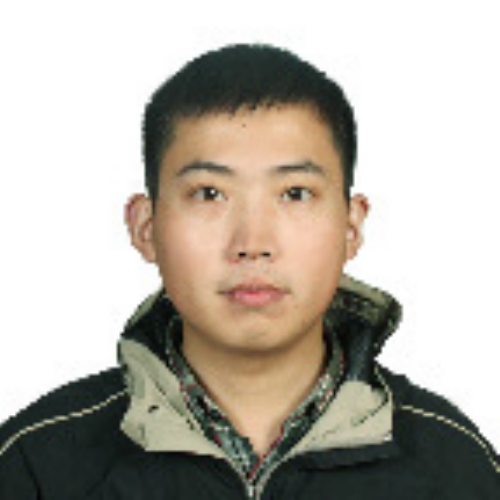}}] {Xiaopeng Li}
received his B.S. degree in Information Engineering from South China University of Technology, Guangzhou, P. R. China, in 2013. He is currently pursuing the M.S. degree in the School of Information and Communication Engineering at Beijing University of Posts and Telecommunications, Beijing, P. R. China. His current research interests include belief propagation and cooperative localization.
\end{IEEEbiography}

\begin{IEEEbiography}[{\includegraphics[width=1in,height=1.25in,clip,keepaspectratio]{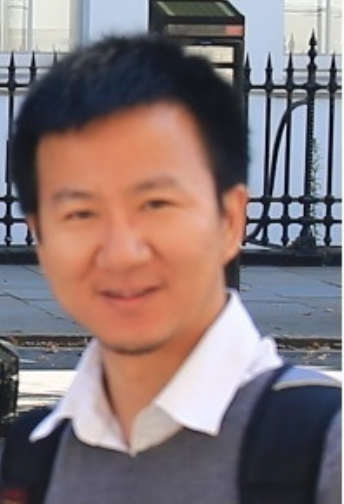}}] {Shaoshi Yang}
(S'09-M'13) received his B.Eng. degree in Information Engineering from Beijing University of Posts and Telecommunications (BUPT), Beijing, China in Jul. 2006, his first Ph.D. degree in Electronics and Electrical Engineering from University of Southampton, U.K. in Dec. 2013, and his second Ph.D. degree in Signal and Information Processing from BUPT in Mar. 2014. He is now working as a Postdoctoral Research Fellow in University of Southampton, U.K. From November 2008 to February 2009, he was an Intern Research Fellow with the Communications Technology Lab (CTL), Intel Labs, Beijing, China, where he focused on Channel Quality Indicator Channel (CQICH) design for mobile WiMAX (802.16m) standard. His research interests include MIMO signal processing, green radio, heterogeneous networks, cross-layer interference management, convex optimization and its applications. He has published in excess of 30 research papers on IEEE journals and conferences. 

Shaoshi has received a number of academic and research awards, including the prestigious Dean's Award for Early Career Research Excellence at University of Southampton, the PMC-Sierra Telecommunications Technology Paper Award at BUPT, the Electronics and Computer Science (ECS) Scholarship of University of Southampton, and the Best PhD Thesis Award of BUPT. He is a member of IEEE/IET, and a junior member of Isaac Newton Institute for Mathematical Sciences, Cambridge University, U.K. He also serves as a TPC member of several major IEEE conferences, including \textit{IEEE ICC, GLOBECOM, PIMRC, ICCVE, HPCC}, and as a Guest Associate Editor of \textit{IEEE Journal on Selected Areas in Communications.} (https://sites.google.com/site/shaoshiyang/) 
\end{IEEEbiography}

\begin{IEEEbiography}[{\includegraphics[width=1in,height=1.25in,clip,keepaspectratio]{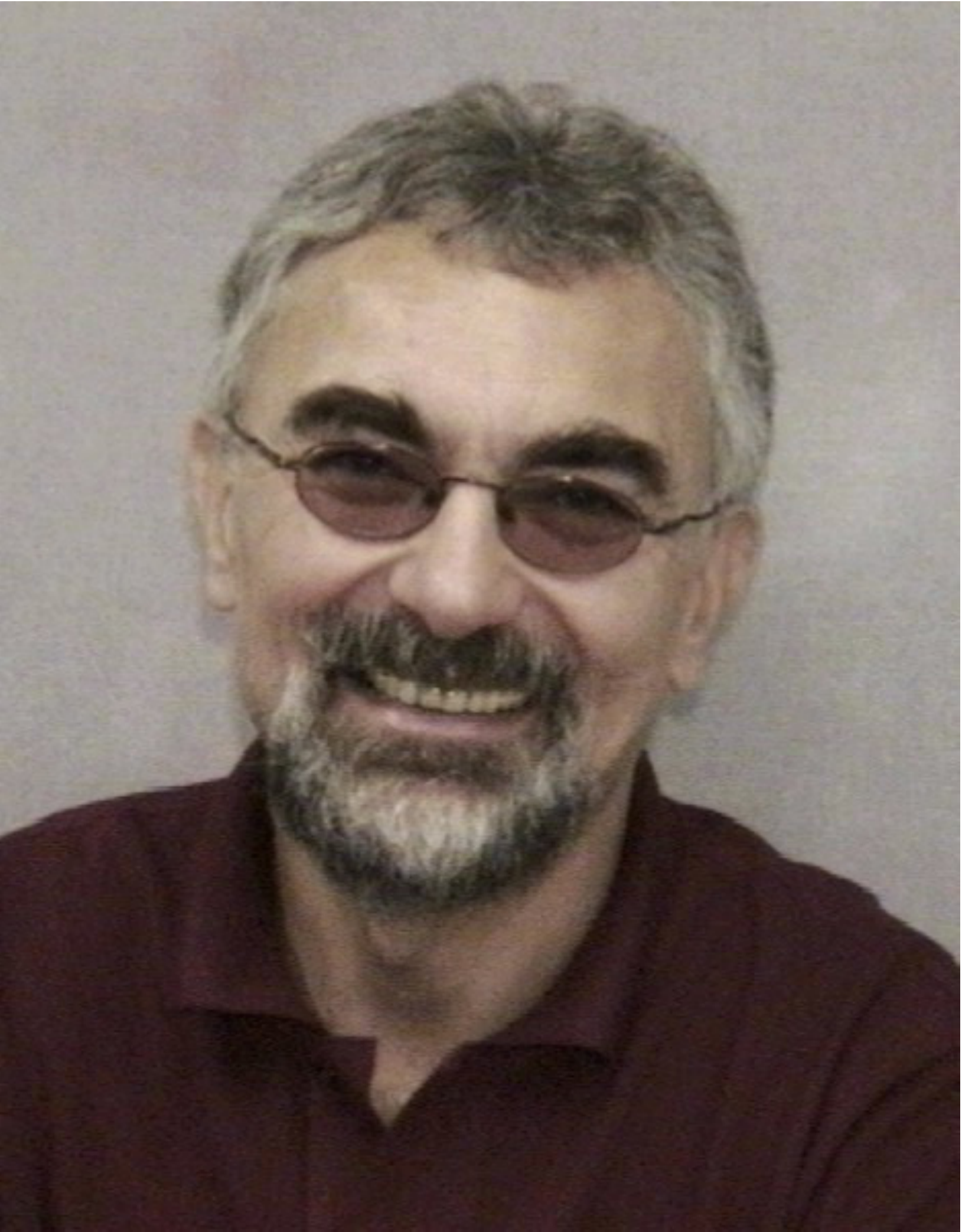}}] {Lajos Hanzo}
(M'91-SM'92-F'04) received his degree in electronics in
1976 and his doctorate in 1983.  In 2009 he was awarded the honorary
doctorate ``Doctor Honoris Causa'' by the Technical University of
Budapest.  During his 39-year career in telecommunications he has held
various research and academic posts in Hungary, Germany and the
UK. Since 1986 he has been with the School of Electronics and Computer
Science, University of Southampton, UK, where he holds the chair in
telecommunications.  He has successfully supervised 100+ PhD students,
co-authored 20 John Wiley/IEEE Press books on mobile radio
communications totalling in excess of 10 000 pages, published 1400+
research entries at IEEE Xplore, acted both as TPC and General Chair
of IEEE conferences, presented keynote lectures and has been awarded a
number of distinctions. Currently he is directing a 60-strong
academic research team, working on a range of research projects in the
field of wireless multimedia communications sponsored by industry, the
Engineering and Physical Sciences Research Council (EPSRC) UK, the
European Research Council's Advanced Fellow Grant and the Royal
Society's Wolfson Research Merit Award.  He is an enthusiastic
supporter of industrial and academic liaison and he offers a range of
industrial courses. 

Lajos is also a Fellow of the Royal Academy of Engineering, of the Institution
of Engineering and Technology (IET), and of the European Association for Signal
Processing (EURASIP). He is a Governor of the IEEE VTS.  During
2008 - 2012 he was the Editor-in-Chief of the IEEE Press and a Chaired
Professor also at Tsinghua University, Beijing. He 
has 22 000+ citations. For further information on research in progress and associated
publications please refer to http://www-mobile.ecs.soton.ac.uk 
\end{IEEEbiography}

\end{document}